\documentclass{emulateapj}
\usepackage{longtable}

\newcommand{\lsun}{\mbox{L$_\odot$}}
\newcommand{\msun}{\mbox{M$_\odot$}}

\newcommand{\mean}[1]{\mbox{$\langle#1\rangle$}} 
\newcommand{\av}{\mbox{$A_V$}} 
\newcommand{\lbol}{\mbox{$L_{bol}$}} 
\newcommand{\tbol}{\mbox{$T_{bol}$}} 

\shorttitle{Protostars in Serpens, Perseus, and Ophiuchus}
\shortauthors{Enoch et al.}

\begin{document}

\title{Properties of the Youngest Protostars in Perseus, Serpens, and Ophiuchus}

\author{Melissa L. Enoch\altaffilmark{1}, Neal J. Evans II\altaffilmark{2}, Anneila I. Sargent\altaffilmark{3}, and Jason Glenn\altaffilmark{4}}

\email{MLE: menoch@astro.berkeley.edu}
\altaffiltext{1}{Department of Astronomy, Univ. of California, Berkeley, CA, 94720}
\altaffiltext{2}{The University of Texas at Austin, Astronomy Department, 1 University Station C1400, Austin, TX, 78712-0259}
\altaffiltext{3}{Division of Physics, Mathematics \& Astronomy, California Institute of Technology, Pasadena, CA 91125}
\altaffiltext{4}{Center for Astrophysics and Space Astronomy, 389-UCB, University of Colorado, Boulder, CO 80309}

\begin{abstract}

We present an unbiased census of deeply embedded protostars in
Perseus, Serpens, and Ophiuchus, assembled  by combining large-scale
1.1~mm Bolocam continuum  and \textit{Spitzer} Legacy surveys.  We
identify protostellar candidates based on  their mid-infrared
properties, correlate their positions with 1.1~mm core positions from
\citet{enoch06}, \citet{young06}, and \citet{enoch07},  and construct
well-sampled SEDs  using our extensive wavelength coverage
($\lambda=1.25-1100~ \micron$).  
Source classification based on the bolometric temperature
yields a total of 39 Class~0 and 89 Class~I sources in the three cloud
sample.  We compare to protostellar evolutionary models using the
bolometric temperature-luminosity diagram, finding a population of low
luminosity Class~I sources that are inconsistent with constant or
monotonically decreasing mass accretion rates.  This result argues
strongly  for episodic accretion during the Class~I phase,  with more
than 50\% of sources in a ``sub-Shu''  ($dM/dt < 10^{-6}$
\msun\ yr$^{-1}$) accretion state.  
Average spectra are compared to protostellar radiative transfer 
models, which match the observed spectra fairly well in Stage~0, but 
predict too much near-IR and too little mid-IR flux in Stage~I.
Finally, the relative number of
Class~0 and Class~I sources are used to estimate the lifetime of the
Class~0 phase;  the three cloud average yields a Class~0 lifetime of
$1.7\pm0.3\times10^5$ yr,  ruling out an extremely rapid early
accretion phase.  Correcting  photometry for extinction results in a
somewhat shorter lifetime ($1.1\times10^5$ yr).  In Ophiuchus,
however, we find very few Class~0 sources
($N_{\mathrm{Class~0}}/N_{\mathrm{Class~I}} \sim 0.1-0.2$), similar to
previous studies of that cloud.   The observations suggest a
consistent picture of nearly constant \textit{average} accretion rate
through the entire embedded phase, with accretion becoming episodic by
at least the Class I stage, and possibly earlier.

\end{abstract}

\section{Introduction}

The problem of how low mass stars like the sun form has been studied
extensively over the last few decades.   Compared to more evolved
protostars and pre-main sequence objects, however, the earliest stages
of the star formation process, from the formation of dense cores
through the main mass accretion phase, are relatively poorly
understood.  This lack of information is due in large part to the
difficulty of observing young, deeply embedded sources, which are
shrouded within dense protostellar envelopes and only observable via
reprocessed emission at mid-infrared to millimeter wavelengths.
Furthermore, most previous observations of deeply embedded objects
have naturally focused on a small number of very bright or well known
sources, due to the sensitivity and resolution limitations of
long-wavelength surveys.  Understanding the formation of typical
stars requires complete samples of young objects, over molecular cloud
scales.

Currently, the details of the early evolution of protostellar sources
are extremely uncertain, including mass accretion rates during the
Class 0 and Class I phases.  In addition, measurements of the
timescales associated with the earliest stages vary considerably,
ranging from $10^5$ to $10^7$~yr for prestellar cores \citep{wt07} and
from $10^4$ to a few $\times10^5$~yr for Class~0 \citep{am94,vrc02}.
In fact, the association of Class~0 and Class~I with distinct
evolutionary stages is still a matter of debate \citep[e.g.,][]{rayj}.
Large surveys at mid-infrared to millimeter wavelengths, where the
SEDs of embedded sources peak, are essential for understanding how
protostars evolve through their earliest  stages.  In addition to
providing complete samples of young sources, large surveys are also
important for characterizing variations in the star formation process
with environment.

With a few notable recent exceptions \citep{jorg07,hatch07}, previous
samples of very young protostars have typically been compiled from many
different surveys, and suffered from systematics, unquantified
environmental effects, and small number statistics.  
We recently completed large-scale 1.1~mm continuum surveys of Perseus,
Serpens, and Ophiuchus with Bolocam at the Caltech Submillimeter
Observatory (CSO).  Maps have a resolution of $31\arcsec$ and cover
7.5 deg$^2$ in Perseus (140~pc$^2$ at our adopted cloud distance of
$d=250$~pc), 10.8 deg$^2$ in Ophiuchus (50~pc$^2$ at $d=125$~pc), 
and 1.5 deg$^2$ in Serpens (30~pc$^2$ at $d=260$~pc) \citep[][hereafter
  Papers I, II, and III, respectively]{enoch06,young06,enoch07}.
These Bolocam surveys complement large \textit{Spitzer}
Space Telescope IRAC and MIPS maps of the same clouds from the ``From
Molecular Cores to Planet-forming Disks'' \textit{Spitzer}
Legacy program (``Cores to Disks'' or c2d; \citealt{evans03}).

Millimeter emission traces the dust in dense starless cores and
protostellar envelopes, and provides a measure of  core and envelope
properties, including sizes, masses, and spatial distribution.
\textit{Spitzer} IRAC and MIPS observations are complementary in that
they provide information about the properties of any young protostars
embedded within dense cores.  
Combining these data enables us to assemble a mass limited,
unbiased census of the youngest star-forming objects in three different
environments, including prestellar cores, Class~0, and Class~I protostars.

In Paper III we looked at how the global cloud environment influences
the properties of star-forming cores \citep{enoch07}.  In a companion
paper to this work \citep{enoch08}, we examine the properties of
prestellar and protostellar cores in Perseus, Serpens, and Ophiuchus,
focusing on the prestellar core mass distribution and the lifetime of
prestellar cores.   A similar analysis comparing the c2d
\textit{Spitzer} data to large SCUBA maps of Perseus and Ophiuchus has
recently been carried out by \citet{jorg07,jorg08}, focusing on the
difference between starless and protostellar cores, as well as cloud
properties such as the star formation efficiency and how it varies
with spatial clustering.  We follow \citet{dif07} in
defining millimeter cores containing a compact luminous internal
source (i.e., an embedded protostar) as ``protostellar cores'',
regardless of whether the final object will be stellar or sub-stellar
in nature.  We use ``starless cores'' to refer to dense cores without
an internal luminosity source, and ``prestellar cores''  as starless
cores that are likely to be gravitationally bound \citep[see][]{enoch08}.  

In this work
we exploit the combined power of millimeter and mid- to far-infrared
observations to study the evolution of young protostars embedded
within the protostellar cores.
Extensive wavelength coverage from $\lambda=1.25-1100~ \micron$ allows us
to trace the evolution of protostellar sources in their main
mass  accretion phase, from formation through the end of the embedded
phase.  We follow \citet{rob06} and \citet{crapsi08} in using
``Stage'' (e.g., Stage 0, I, II, III) to refer to a source's true
physical nature, regardless of its observed properties, while the
corresponding ``Class'' refers to the observational classification,
typically based on the near- to mid-infrared slope $\alpha_{IR}$
\citep{als87,greene94} or on the bolometric temperature \citep{ml93}.

We assume that Stage~0 and Stage~I refer to an evolutionary sequence
of embedded protostars with $M_{*} < M_{env}$ and $M_* > M_{env}$,
respectively \citep[e.g.][]{andre94}, where $M_{*}$ is the protostar
mass and $M_{env}$ the envelope mass.  Similarly, we assume Stage~II
refers to pre-main sequence stars with very little remaining envelope
($M_{env} < 0.1 M_{\sun}$; \citealt{crapsi08}).  
Our adopted definitions of various Classes and Stages are summarized in Table~\ref{deftab}.
Ideally, Class~I would directly correspond to Stage~I, etc., 
but as shown by \citet{rob06} and \citet{crapsi08}, geometric effects 
can cause, for example, Stage~II sources to be classified as Class~I.

In \S\ref{datasec} we describe the 1.1~mm and \textit{Spitzer}
infrared (IR) data, including how protostellar candidates are
identified  and their association with 1.1~mm cores.  We calculate bolometric
luminosities and temperatures, envelope masses, and discuss completeness
(\S\ref{embprops}).  Protostellar classification methods are compared
in \S\ref{classsec}.  Spectral characteristics of Class~0, I and II
sources detected at 1.1~mm are discussed in \S\ref{charsec}, including
selected individual sources (\S\ref{indsec}). In \S\ref{otherclasssec}
we explore alternative classification schemes made possible by our
large sample, and compare the average observed spectra to
protostellar models in \S\ref{modsec}.  Sources are placed on a
bolometric temperature-luminosity diagram for
comparison to protostellar evolutionary models and to study their
luminosity evolution, mass accretion rates, and envelope evolution (\S\ref{evolsec}).
Finally, in \S\ref{lifesec} we calculate lifetimes for the Class~0
phase.

\section{Combining Bolocam and Spitzer c2d Data}\label{datasec}

Both Bolocam 1.1~mm and \textit{Spitzer} maps were designed to cover down to
a visual extinction of $\av \gtrsim 2$~mag in Perseus, $\av \gtrsim
3$~mag in Ophiuchus, and $\av \gtrsim6$~mag in Serpens
\citep{evans03}.  
The actual overlap in area between Bolocam and IRAC maps is
shown in  Figure~1 of Papers I, II, and III for Perseus, Ophiuchus, and
Serpens, respectively.  Catalogs
listing c2d measured \textit{Spitzer}  fluxes of all sources in each
of the three clouds, as well as near-infrared fluxes for sources in
the 2MASS catalogs, are available through the  \textit{Spitzer}
database \citep{evans07}.  We utilize wavelength coverage from
$\lambda=1.25$ to $1100~ \micron$, using 2MASS
(1.25, 1.65, 2.17~$\micron$), IRAC (3.6, 4.5, 5.8, 8.0~$\micron$), MIPS
(24, 70, 160~$\micron$), and Bolocam (1.1~mm)  data.  
Note that $160~ \micron$ flux measurements are not included in 
the c2d delivery catalogs due to substantial uncertainties and 
incompleteness, but are included here when possible.
Photometry at $160\micron$ is discussed in \citet{reb07} and 
\citet{harv07a}, where $160\micron$ fluxes for point sources 
in Perseus and Serpens are also given. 

\begin{figure*}[!ht]
\vspace{0.4in}
\begin{center}
\includegraphics[width=6.2in]{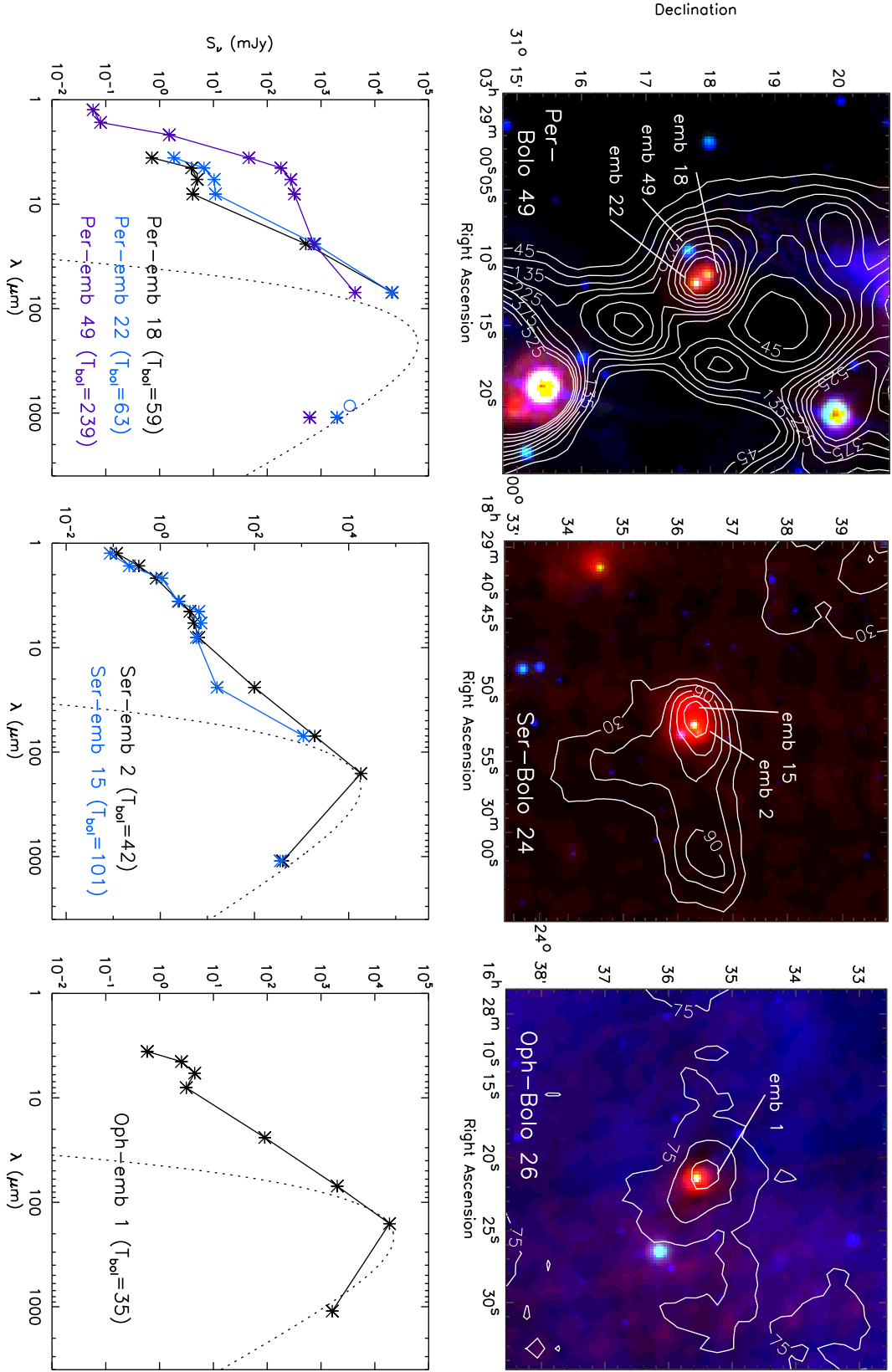}
\caption{Three-color Spitzer images $8~\micron$ 
(blue), $24~\micron$ (green), and $70~ \micron$ (red) of
  selected embedded protostar candidates  in Perseus, Serpens, and
  Ophiuchus, with Bolocam 1.1~mm contours.  
The Bolocam ID of the centered core and embedded
  protostar IDs from  Tables~\ref{boltabp}--\ref{boltabo} given, and
  contour intervals for the 1.1~mm emission are
  $3,6...15,20,25,...\sigma$.  
Spectral energy distributions
  (SEDs) of the candidates are plotted in the lower panels.  SEDs
  include 2MASS, IRAC, and MIPS photometry from the c2d database and
  1.1~mm fluxes (asterisks). Published $350~ \micron$ and $850~
  \micron$ points (open circles) are also shown when available.   A
  modified blackbody spectrum ($T=15$~K, $\beta=1$) is overlaid for
  reference.  
\label{psp3color}}
\end{center}
\end{figure*}

Basic data papers describe the processing and analysis of the
\textit{Spitzer} IRAC and MIPS maps of Perseus, Serpens,  and
Ophiuchus, as well as presenting general properties of the sources in
each cloud, such as color-color and color-magnitude diagrams
\citep{jorg06,harv06, reb07,harv07a}. The young stellar object (YSO) 
population in Serpens is discussed in detail by \citet{harv07b}.   
Here we are most interested
in the young protostellar sources that are most likely to be embedded
in the millimeter cores detected with Bolocam.  For the following we
will use the term ``embedded protostar candidate'' in general to
encompass candidate Stage~I and younger objects.  Although it will
become apparent in \S\ref{pidsect} that our criteria also pick up a
number of Stage~II sources, in general we focus on sources with
evidence for a protostellar envelope (Stage~0 and I). 

Figure~\ref{psp3color} shows the result of combining \textit{Spitzer}
and Bolocam data, for a few examples of embedded protostellar sources
in each cloud.  Images are three-color \textit{Spitzer} maps (8.0, 24, 70 $\micron$), 
with 1.1~mm contours overlaid.  The Bolocam ID of the
associated 1.1~mm core is given in each panel, as well as the embedded
source IDs from Tables~\ref{boltabp}--\ref{boltabo}.   We use the
2MASS, IRAC, MIPS, and Bolocam data to construct complete 
SEDs from $2$ to $1100~ \micron$ for each
candidate protostellar source, shown in the lower panels (see also \S\ref{embprops}).
Additional SHARC~II $350~ \micron$ fluxes \citep{wu07} and SCUBA 
$850~ \micron$ fluxes \citep[][open circles]{kirk06} are included when
available.  Modified black-body ($S_{\nu} \propto\nu^{\beta} B_{\nu}(T)$) 
curves for a temperature of 15~K and $\beta=1$ are shown
for reference in Figure~\ref{psp3color} (dotted lines).

Protostellar sources in our sample may be isolated (e.g., Per-Bolo~57)
or lie in crowded regions (e.g., Per-Bolo~49).
Approximately 20\%--50\% of the time, more
than one protostellar source lies within a single 1.1~mm core 
(20\% of the embedded protostar sample in Ophiuchus, 40\% in Perseus, 
and 55\% in Serpens).  
Much of this difference is certainly due to the lower resolution of Bolocam (31\arcsec)
compared to  \textit{Spitzer} (7\arcsec at $24~ \micron$),
causing nearby \textit{Spitzer} sources to be blended in the 1.1~mm
map, although the envelopes detected at 1.1~mm are also physically more 
extended than the region emitting at \textit{Spitzer} wavelengths. 
Sometimes the SEDs of such multiple  sources look similar to
each other (Per-emb~18 and 22 in the lower-left  panel of
Figure~\ref{psp3color}), and sometimes quite different  (Per-emb~49
of the same panel).

Although the coverage of the IRAC, MIPS and Bolocam maps overlaps
nearly perfectly for our purposes, there are a few cases in which
embedded protostar candidates are outside the boundaries of the MIPS
$70~ \micron$ map (three sources in Serpens) or 1.1~mm map (one source
in Perseus).   In addition, a number of bright sources in each cloud
are saturated in the $24\micron$ or $70\micron$ c2d maps.  In these
cases we substitute IRAS $25\micron$ or $60\micron$ fluxes is the
source does not appear to be blended in a visual inspection of the the
IRAS maps.  

The $160~ \micron$ maps are often saturated
near bright sources and in regions of bright extended emission, such
as near bright clusters of sources.  Reliable $160~ \micron$ fluxes are
especially difficult to determine in crowded regions, due both to the
large beam size ($40\arcsec$) and to saturation issues.  The lack of
$160~ \micron$ data is most problematic in Ophiuchus,  where the
$160~ \micron$ maps are saturated in all of the dense source regions.
Even for isolated sources, the measured $160~ \micron$ flux density,
determined from a point spread function (PSF) fit, may be
underestimated if the source is extended.  The effects of these issues
on our analysis are discussed in more detail in the appendix
(\S~\ref{appendsect}).

\subsection{Identifying Embedded Protostar Candidates}\label{pidsect}

We form a sample of candidate embedded protostars from the c2d catalogs;
the first cut is based on the source ``class.''   All sources in the
c2d catalogs are assigned a class parameter  based on colors,
magnitudes, and stellar SED fits (see the c2d Delivery
Document \citep{evans07} and \citet{harv07b}).   Class parameters
include ``star'',  ``star+disk'', ``YSOc''  (young stellar object
candidate), ``red'',  ``rising'', ``Galc'' (galaxy candidate), etc.
Embedded protostars will generally be a
subset of YSOc sources, but some  of the most embedded may also be
assigned to the ``red'' class if  they are not detected in all 
IRAC bands.  

Thus, we begin by selecting all sources from the c2d
database that are  classified as ``YSOc'' or ``red.''  From this list,
we keep sources that meet \textit{all} of the following criteria: \\
(a) flux density at $24~ \micron$ $(S_{24\micron}) \ge 3$~mJy \\ 
(b) $S_{24\micron} \ge 5 \alpha_{IR} + 8$~mJy, where $\alpha_{IR}$ is
the  near- to mid-infrared spectral index, determined by a
least-squares fit to photometry between $2$ and $24~ \micron$.  This criteria is 
motivated by a comparison to the carefully vetted Serpens YSO sample of \citet{harv07b}.  \\  
(c) $\nu S_{24\micron} > \nu S_{8\micron}$, i.e., 
the SED is rising from 8 to 24~$\micron$ in $\nu S_{\nu}$ space. \\
(d) $S_{24\micron}$ must be of high quality, i.e., signal to noise (S/N) greater than 7. \\
(e) $S_{24\micron}$ is not a ``band-filled''
flux.   For sources not originally detected in all \textit{Spitzer}
bands, a  flux or upper limit is measured at the source position
(band-filling, \citealt{harv07b}).   Because the resolution is lower
at $24~ \micron$ than at the shorter  wavelengths, some IRAC-only
sources have unreliable band-filled fluxes at $24~ \micron$
(e.g., sources are confused with the PSF wings of a nearby source).

In addition to sources that meet the above criteria,  we include 
any $70~ \micron$ point sources not classified as
galaxy candidates (``Galc'').   Note that these $70 \micron$ sources
need not be classified  as ``red'' or ``YSOc''.  In each cloud, a
number  of deeply embedded sources that are bright at $70~ \micron$
but very weak at $24~ \micron$ (e.g., HH211 in Perseus) are recovered
by this last criteria (5 in Perseus, 4 in Serpens, and 3 in Ophiuchus), as
are a few very bright sources that are saturated at $24~ \micron$ (6
in Perseus, 2 in Serpens, and 6 in Ophiuchus; these are often
classified as ``rising'').

\begin{figure}[!ht]
\hspace{-0.3in}
\includegraphics[width=3.7in]{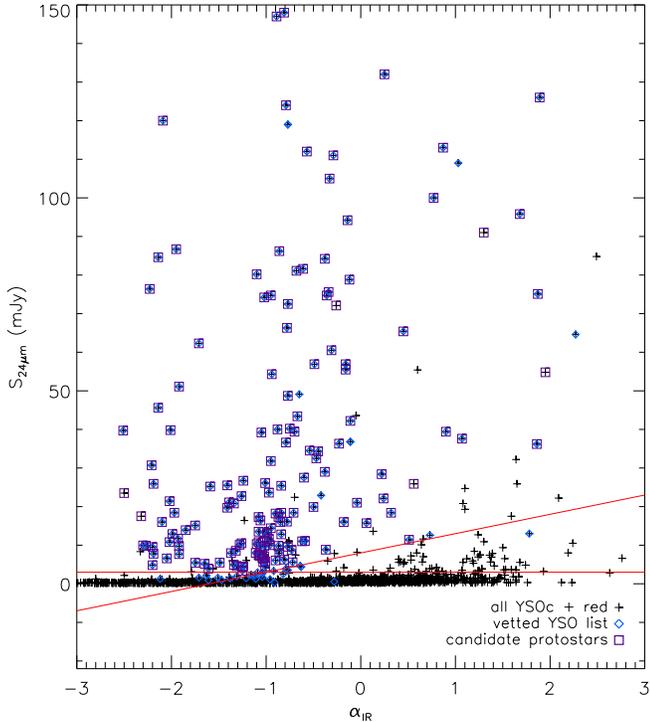}
\caption
{Plot of $S_{24\micron}$ versus spectral index $\alpha_{IR}$ for
  sources in Serpens, to demonstrate selection criteria for the
  embedded protostar candidate samples.  Plus symbols (``+'') indicate
  the sample from which the candidate embedded protostars is drawn,
  including all catalog sources labeled as ``YSOc'' or ``red'' as well
  as non-''Galc'' $70\micron$ sources.   Lines  show $24~ \micron$ flux cuts
  imposed by criteria (a) and (b), which remove the majority of
  spurious or background sources.  Boxes indicate the final embedded
  protostar sample, after applying criteria (a)-(d).  Smaller blue
  diamonds show the carefully vetted YSO sample from \citet{harv06}.
  The two samples agree fairly well, but because we select for
  embedded sources we miss many of the more evolved YSOs in the vetted
  YSO list (Class II/III).  There are a few sources in the embedded
  protostar sample not in the \citet{harv06} list, two of which are
  likely to be real (see text), while the others are rejected when
  examined by eye.
\label{selectfig}}
\end{figure}

Figure~\ref{selectfig} plots $S_{24\micron}$ versus $\alpha_{IR}$ for
embedded candidates in Serpens, where ``+'' symbols indicate the
original (``YSOc''+``red''+ $70\micron$ sources) sample for Serpens, and
boxes indicate our embedded protostar candidates  after applying
criteria (a)--(e).  For comparison, diamonds indicate the carefully vetted  
Serpens YSO list from \citet{harv07b}. The majority of ``+''-only points, which were
rejected as true YSOs by \citet{harv07b}, are removed by criteria (a)
and (b) (shown as solid lines).  More than half of the vetted YSOs do
not appear in the candidate embedded protostar sample 
(diamond, no box); the majority of these are rejected by the rising SED
criteria (c) and as they are primarily 
classified as ``star+disk'' we do not expected them to be embedded.
There are a few sources in our sample that
are not in the vetted YSO sample (box, no diamond).  A few
of these were easily identified as non-protostellar, and rejected,
when examining by eye.  We do identify three embedded protostar
candidates in Serpens that are not in the vetted YSO list but seem to be
associated with  1.1~mm emission (see \S\ref{assocsec}): one ``red'',
and two ``rising''.

\begin{figure}[!ht]
\hspace{-0.2in}
\includegraphics[width=3.6in]{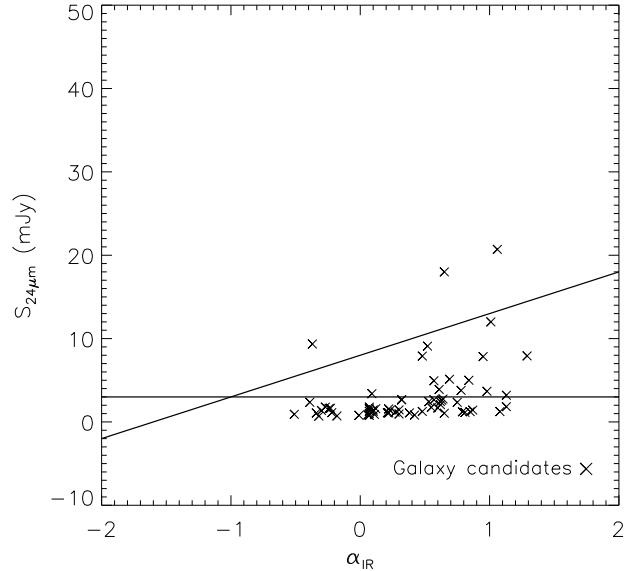}
\caption[Protostellar selection criteria applied to galaxy candidates]
{Plot of $S_{24\micron}$ versus spectral index $\alpha_{IR}$ for
galaxy candidates (``Galc'') in Serpens.  Selection criteria (a) and
(b) are overlaid, as in Figure~\ref{selectfig}.  These selection criteria
eliminate most of the parameter space inhabited by extragalactic
sources.
\label{selectfig2}}
\end{figure}

We include red sources to ensure that we identify the most embedded
protostars,  but we need to reject the large percentage of these
sources that are likely to be galaxies.   As shown in
Figure~\ref{selectfig2}, criteria (a) and (b) are efficient at
eliminating  extragalactic sources.  Here $S_{24\micron}$ is plotted
versus $\alpha_{IR}$ for all ``Galc'' galaxy candidates in Serpens.
As in Figure~\ref{selectfig}, lines show criteria (a) and (b)

After forming our embedded candidate samples based on the above
criteria, the images and SEDs of each source are examined by eye to
remove any galaxies that are extended in the near-infrared, and other
obviously non-embedded sources.  SEDs include 2MASS, IRAC, MIPS
and 1.1~mm fluxes.   In some cases, there is no available point-source
flux at 70 or $160~ \micron$ even if there is emission at the position
of the source, generally because the source is extended at these
wavelengths.  In these cases a flux density is measured by hand, if
possible, using aperture photometry.  
Large uncertainties (50\% or more) are associated with these 
band-filled flux measurements.

\subsection{Association with a 1.1~mm Core}\label{assocsec}

The next step after assembling a sample of candidate embedded
protostars is to determine which are associated with 1.1~mm
emission.  The correlation between candidate protostar and 1.1~mm core
positions was done in \citet[][see \S2.3 and
 Figure~2]{enoch08}, following a similar analysis by \citet{jorg07}; 
we summarize the results here.  We found 
that the coldest protostellar candidates ($T_{bol} <
300$~K, see \S\ref{bolsec})  are primarily located within $1.0\times
\theta_{1mm}$ of a millimeter core position, where $\theta_{1mm}$ is
the angular  FWHM size of the 1.1~mm core, as expected if they are
deeply embedded.  Based on that analysis, a given embedded protostar
candidate is assumed to be associated with a millimeter core if it is
located within $1.0\times \theta_{1mm}$ of the 1.1~mm core centroid
position.

We also found in \citet{enoch08}, based on analysis of a spatially
random distribution of sources, that we can  expect approximately 5
false associations with 1.1~mm cores in each cloud.  Using a more
restrictive criteria (e.g., $0.5\times \theta_{1mm}$) would reduce the
number of false associations, but would likely miss at least a few
embedded protostars.  For reference, adopting $0.5\times
\theta_{1mm}$ would result in 10 fewer embedded protostars in Perseus,
7 in Serpens, and 16 in Ophiuchus.

If an embedded protostar candidate is found to be associated  with a
millimeter core by the above criteria, the 1.1~mm flux is included in
the protostellar SED, and is used to calculate  an envelope mass
($M_{env}$, \S\ref{envmasssec}).  If a candidate
is not located within $1.0\times \theta_{1mm}$ of a millimeter core
position, we calculate a flux or upper limit from the original 1.1~mm
map (although sources with upper limits at 1.1~mm are not included in the 
final lists of embedded protostars given in Tables~\ref{boltabp}--\ref{boltabo}). 

In a number of cases  (10 in Perseus, 3 in Serpens, and 11 in
Ophiuchus) there is clearly 1.1~mm flux at the protostar position that
was not identified as a core in the original 1.1~mm source extraction
because it is below $5\sigma$ (where $\sigma$  is the local rms noise,
see Paper~I) or is in a confused region of the map.   If this emission
exceeds  $3\sigma$, we measure a ``band-filled'' 1.1~mm flux using a
small aperture (30\arcsec--40\arcsec).  
We also re-compute the 1.1~mm
flux density in small apertures for all sources in regions of blended
1.1~mm emission.  When more than one protostar candidate is associated
with a single compact millimeter core, we divide the total 1.1~mm flux
of the core equally between the protostellar sources.

The approach described above means that the 1.1~mm flux is not
measured in a fixed aperture for every source.  In crowded regions or
for sources with close neighbors, the 1.1~mm flux is measured in a
30\arcsec--40\arcsec diameter aperture centered on the
\textit{Spitzer} source position.  These sources are indicated by a
footnote in Tables~\ref{boltabp}--\ref{boltabo}.  For other sources,
the Bolocam 1.1~mm core flux from Papers I--III is used.  Typically, but not
always, the total core flux is used here (integrated in the largest aperture, 
from $30\arcsec-120\arcsec$
diameters in steps of $10\arcsec$, that is smaller than the distance to the 
nearest neighboring source); fluxes in 
30, 40, 80, and 120\arcsec\ apertures can also be found in Papers I--III.

Approximately 50\% of the embedded protostar candidates in each cloud
are lacking 1.1~mm emission, even after  re-examining the 1.1~mm maps
at each source position.  Most of these sources appear to be Class~II
objects, with little or no remaining envelope.  Any 1.1~mm emission
from these objects, therefore, is likely below our detection limit of
$\sim 0.1 M_{\sun}$ (\S~\ref{completesec}).  A few of these  may actually be 
low luminosity
sources that are truly embedded, with low mass envelopes  below our
detection limit. For this reason, we only claim completeness to
embedded protostars with $M_{env} > 0.1 ~ \msun$.
Note that this corresponds 
well to the \citet{crapsi08} definition of Stage~I.

\section{Properties of Embedded Protostars}\label{embprops}

\subsection{Bolometric Luminosity and Temperature}\label{bolsec}

Our extensive wavelength coverage allows us to construct well-sampled
SEDs for the embedded protostar candidates in all three clouds, from which
we calculate a bolometric luminosity ($L_{bol}$)  and bolometric
temperature ($T_{bol}$) for each source.  The bolometric luminosity is
calculated by integrating the SED ($S_{\nu}$)  over frequency:
\begin{equation}
L_{bol} = 4 \pi d^2 \int S_{\nu} d\nu.   \label{lboleq}
\end{equation}
The bolometric temperature is defined as the temperature of a
blackbody  with the same mean frequency as the source SED, and is
given by 
\begin{equation}
T_{bol} = 1.25 \times 10^{-11} \mean{\nu}\mathrm{~K }
\end{equation}
\citep{ml93}, where the mean frequency is
\begin{equation}
\mean{\nu} = \frac{\int \nu S_{\nu} d\nu}{\int S_{\nu} d\nu}. \label{meanfreqeq}
\end{equation}
Two methods for approximating the integrations over frequency 
for finitely sampled SEDs (midpoint and prismoidal) are discussed in the appendix
(\S\ref{appendsect}).

Tables~\ref{boltabp}--\ref{boltabo} list the  bolometric temperatures
and luminosities derived for all embedded protostars in each cloud.
As we are primarily interested in young, embedded objects, only
sources  with detectable 1.1~mm emission, which likely have
substantial envelopes,  are included.  Sources are listed by
increasing $T_{bol}$, and identified as, e.g., ``Per-emb\#'', as well
as by their c2d name (SSTc2dJ...), which also gives the position.
When sources are saturated at $24$ or $70\micron$, the IRAS $25$ or
$60\micron$ flux is utilized when not affected by blending.  In these
cases a note is made in Tables~\ref{boltabp}--\ref{boltabo}.  The
correction of saturated fluxes can increase the luminosity by more
than a factor of two; for saturated sources with no reliable IRAS flux
(also noted in Tables~\ref{boltabp}--\ref{boltabo}), the luminosity
will be an underestimate.

All $T_{bol}$ and $L_{bol}$ values quoted use the midpoint integration method.  
The difference between the values calculated by the midpoint and prismoidal
integration methods (given in parentheses in Tables~\ref{boltabp}--\ref{boltabo}) 
gives a more realistic measure of the uncertainties in $T_{bol}$ and $L_{bol}$ 
than the formal fitting errors (which are typically 10\%).
Uncertainties can be larger than 50\% depending on whether or
not a $160~ \micron$ flux is available, and there is an additional systematic 
uncertainty of 15--25\% from finite sampling errors (\S\ref{appendsect}).  

Systematic errors introduced
by missing $160~ \micron$ fluxes are investigated in
\S\ref{appendsect}.  Without a $160~ \micron$ measurement,  $T_{bol}$
will almost certainly be an overestimate for very cold sources, which
may bias our classification of protostellar candidates (\S\ref{classsec}).
Ophiuchus will be most affected, as only four sources in that cloud
have reliable $160~ \micron$ fluxes.

\subsubsection{Correcting for Extinction}\label{deredsec}

One might argue that we should correct our photometry for extinction
before calculating $L_{bol}$ and $T_{bol}$.  The effects of extinction
are typically ignored for Class~0 and Class~I sources, for which one
might expect the foreground extinction to be negligible relative to
the effect of the envelope itself.  While these observed values are
most easily compared to the majority of previous work, \citet{chen95}, who connected
$T_{bol}$ to the classes defined by $\alpha$, did correct the observed
flux densities for extinction before computing $T_{bol}$.  Therefore,
we compute extinction corrected values for comparison (see \citealt{evans08}
for a more complete discussion of the
effects of dereddening).  We use the mean extinction to all Class~II
sources in each cloud ($A_V=5.92$ mag in Perseus, 9.57~mag in Serpens,
an 9.76~mag in Ophiuchus; \citealt{evans08}) to
deredden the photometry of Class~0 and Class~I sources.  

As dereddening has a greater effect on shorter wavelength photometry,
correcting for extinction tends to increase both $T_{bol}$ and
$L_{bol}$.  Throughout this paper we will primarily use observed
values, and \textit{not} extinction corrected values, but the 
effects on derived values such as the Class~0 lifetime (\S~\ref{lifesec}) will be noted.
When used, the extinction corrected bolometric luminosity and
temperature are indicated by $T'_{bol}$ and $L'_{bol}$, respectively.

\subsection{Envelope Mass}\label{envmasssec}

The envelope mass of candidate protostars, $M_{env}$, is calculated
from the flux density at 1.1~mm, $S_{1mm}$: 
\begin{equation}
M = \frac{d^2 S_{1mm}}{B_{1mm}(T_D) \kappa_{1mm}}, \label{masseq}
\end{equation}
where $d$ is the cloud distance, $\kappa_{1mm} =
0.0114$~cm$^2$~g$^{-1}$ is  the dust opacity per gram of gas at
1.1~mm, and $B_{1mm}$ is the Planck function at a dust  temperature of
$T_D$.  The opacity is interpolated from Table~1 column~5 of
\citet{oh94} for dust grains with thin ice mantles. 

\begin{figure*}[!ht]
\begin{center}
\includegraphics[width=6.in]{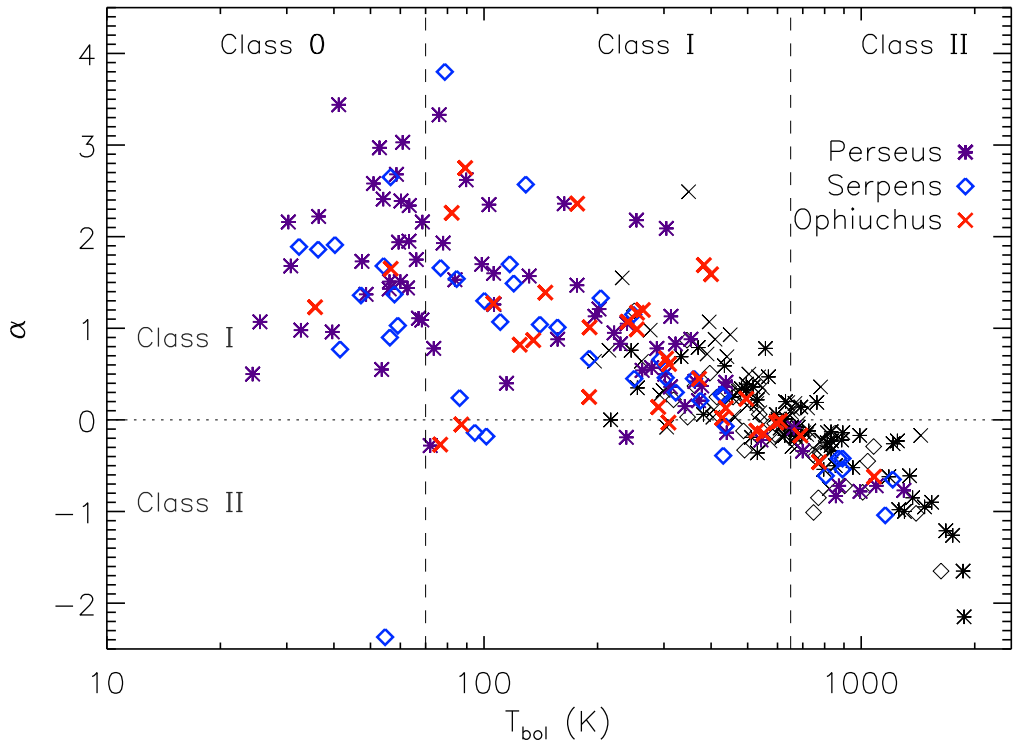} 
\caption
{Near- to mid-IR spectral index ($\alpha_{IR}$) versus bolometric
  temperature ($T_{bol}$)  for the candidate embedded protostar
  samples in Perseus, Serpens, and Ophiuchus.  Bold symbols indicate
  sources that are associated with 1.1~mm emission, and thin symbols
  denote those with upper limits at 1.1~mm.  Standard class divisions
  for both $T_{bol}$ and $\alpha_{IR}$ are shown.  The two methods
  agree fairly well for Class~II and ``warmer'' Class~I sources, but
  very cold ($T_{bol} \lesssim 100$~K) sources have a large range of
  $\alpha_{IR}$ values.  }\label{tbolalpha}
\end{center}
\end{figure*}

We assume a dust temperature of $T_D=15$~K for  protostellar
envelopes, consistent with  average isothermal dust temperatures found
from radiative transfer models of a sample of Class 0 and Class I
protostars \citep{shir02,young03}.  The isothermal dust  temperature
is the temperature that, when used in an isothermal mass equation
(e.g., Eq.~(\ref{masseq}), above) yields the same mass as does a
detailed radiative transfer model that accounts for temperature
gradients.  Dust temperatures will be higher than 15~K close to the
protostar, but the majority of the envelope mass is in the outer,
cooler, regions of the envelope.  
A dust temperature of $10$~K would result in an increase in masses 
by a factor of 1.9, while $T_D=20$~K would decrease masses by a factor of 1.5.

Envelope masses of embedded protostars are listed in
Tables~\ref{boltabp}--\ref{boltabo}.  If the source is associated with
a distinct core,  the Bolocam identification from Papers I--III is
given in the last column.  If the source is not associated with one of
the originally identified cores, but rather the flux has been
``band-filled'' at  1.1~mm, then no Bolocam ID is given.
In these ``band-filled'' cases, the measured 1.1~mm flux can be easily 
re-constructed from the mass.

\subsection{Completeness}\label{completesec}

Because we require a detection at 1.1~mm to be
included in the final source lists, we are clearly incomplete to
objects more evolved than Stage~I, which do not have a substantial
protostellar envelope.  We do detect a few Class~II sources at 1.1~mm; these are discussed in
\S\ref{class2sec}.    
The point-source detection limits of our 1.1~mm surveys limit our
sensitivity to Class~I  sources with $M_{env} \gtrsim 0.09 ~ \msun$ in
Perseus, $M_{env} \gtrsim 0.07 ~ \msun$ in Serpens, and $M_{env}
\gtrsim 0.04 ~ \msun$ in Ophiuchus.  These are $5\sigma$ point source 
detection limits; for very extended envelopes the completeness limits 
will be higher.  For simplicity we take a detection limit of $0.1 \msun$ for all clouds.

As the 1.1~mm detection requirement is more
restrictive than our $24~ \micron$ flux criteria,  we explore the
possibility that we are  missing some low luminosity embedded sources
that are below our 1.1~mm detection threshold.

Taking the 1.1~mm $5\sigma$ detection limits for each cloud (75~mJy in Perseus,
50~mJy in Serpens, and 110~mJy in Ophiuchus), and assuming the
spectrum of a modified blackbody, $S_{\nu} = \nu^{\beta} B_{\nu}(T_D)$,
with $T_D=15$~K, we can estimate the minimum detectable bolometric
luminosity in each cloud.  For $\beta=1$, the minimum $L_{bol}$ is
0.02~\lsun\ in Perseus, 0.01~\lsun\ in Serpens, and 0.01~\lsun\ in
Ophiuchus.  Assuming $T_D=20$~K lowers these values by approximately a
factor of 2, while taking $\beta=2$ increases them by a factor of
4.  Although these are very rough estimates, they agree fairly well
with the lowest observed bolometric luminosities for sources with
1.1~mm detections (0.04~\lsun\ in Perseus, 0.05~\lsun\ in Serpens, and
0.01~\lsun\ in Ophiuchus).  \citet{dunham08} and \citet{harv07a}
demonstrate that the   \textit{Spitzer} c2d surveys are complete  to
young objects with luminosities as low as $0.05 \lsun$.

Finally, a comparison of our source list to the dedicated search for
very low luminosity protostars by \citet{dunham08} confirms that we
are not missing any embedded protostellar sources in Perseus down to
the completeness limits of that survey ($L_{bol} \gtrsim 0.03-0.05
\lsun$).

\section{Source Classification}\label{classsec}

To study the early evolution of protostars, it is necessary to
identify  the evolutionary state of the embedded candidates in our
sample.  This is typically accomplished by classifying sources into
discrete groups based on SED characteristics.  A number of
classification methods are employed in the literature; most often used
are the near- to mid-infrared spectral index  $\alpha_{IR} =
d\log(\lambda F_{\lambda}) / d\log(\lambda)$ \citep{lada87}, the
bolometric temperature $T_{bol}$ (\citealt{ml93}, see \S\ref{bolsec}),
and the ratio of submillimeter to bolometric luminosity
$L_{submm}/L_{bol}$ \citep{awb93}.  Generally, $L_{submm}$ is taken to
be the integrated luminosity at wavelengths $\lambda \ge 350~ \micron$.


\begin{figure*}[!ht]
\begin{center}
\includegraphics[width=5.6in]{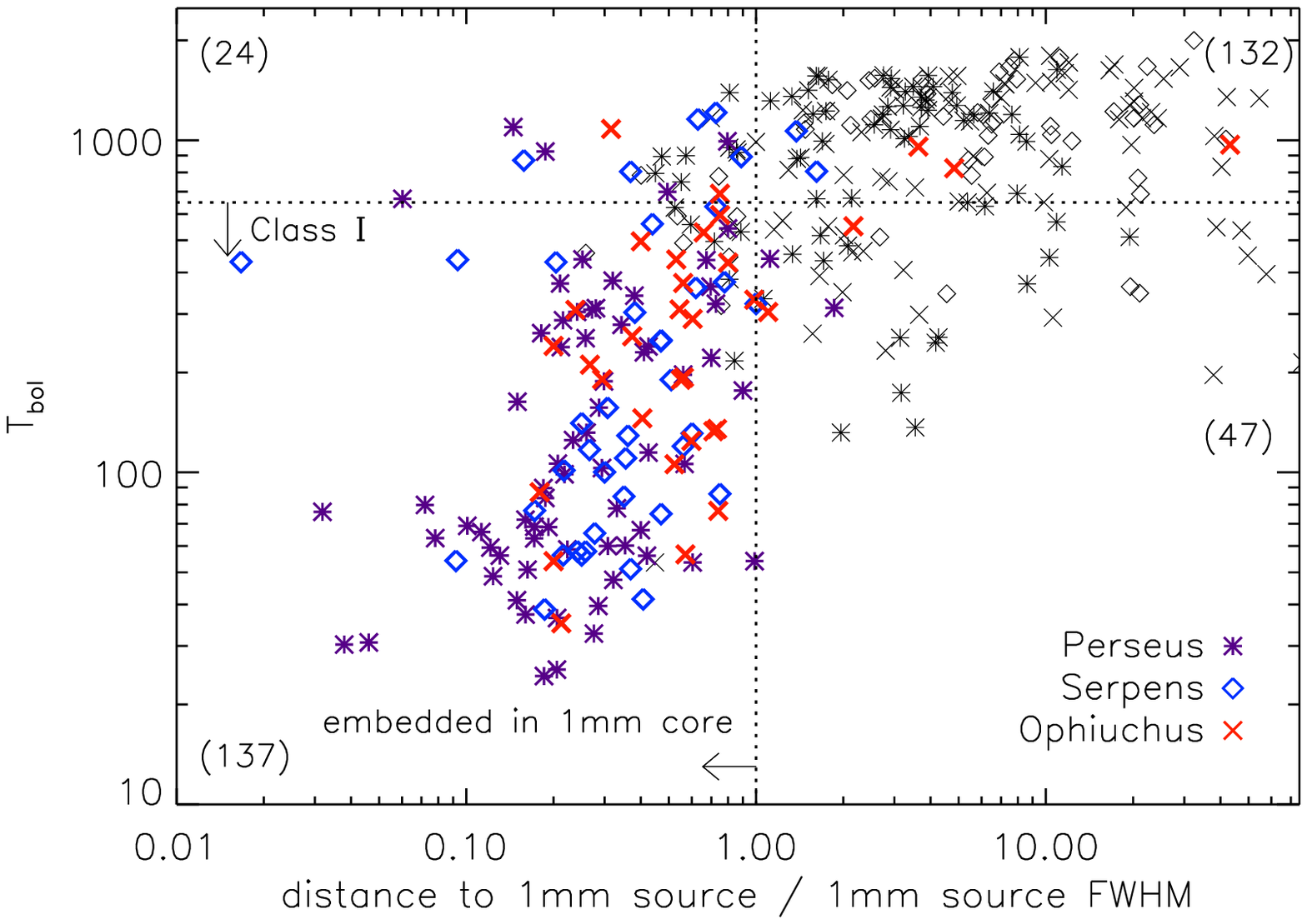} 
\caption
{Distribution of \tbol\  as a function of the distance  to the nearest
  1.1~mm core, for candidate embedded protostars in all three clouds.
  Distances are in units of the 1.1~mm core FWHM size
  ($\theta_{1mm}$), and symbols are as  in Figure~\ref{tbolalpha}.
  Sources within $1.0 \times \theta_{1mm}$ of a 1.1~mm core position
  are considered embedded within that core.  There is a clear
  correlation between smaller distances, or more embedded  sources,
  and lower $T_{bol}$ values, suggesting that $T_{bol}$ is a good
  measure of evolutionary state for deeply embedded sources.
For reference, the total numbers of sources in each quadrant of the plot are given in parentheses.
}\label{distfig}
\end{center}
\end{figure*}

\begin{figure*}[p]
\begin{center}
\includegraphics[width=5.6in]{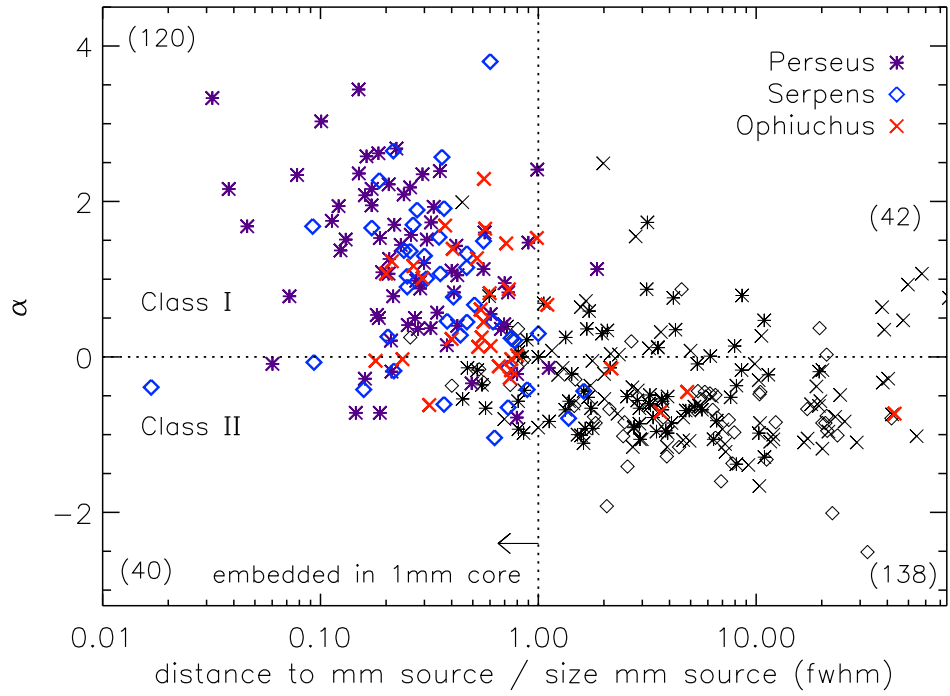} 
\caption{Distribution of $\alpha_{IR}$  as a function of the distance
  to the nearest 1.1~mm core for embedded protostar  candidates in all
  three clouds, similar to Figure~\ref{distfig}.  While there is some
  correlation between higher $\alpha_{IR}$ values and smaller
  distances, it is not as compelling as the correlation observed for $T_{bol}$
  (Figure~\ref{distfig}).  
For reference, the total numbers of sources in each quadrant are given in parentheses.
}\label{adistfig}
\end{center}
\end{figure*}

\clearpage 

Any of these methods must come with the caveat that protostellar mass 
may affect the classification, which only detailed modeling
can resolve \citep[e.g.][]{hatch07}.  We typically have only one flux
measurement for $\lambda >160~\micron$, so an accurate determination
of $L_{submm}$ is not feasible with these data.\footnote{It is possible to calculate
  $L_{submm}$ by assuming a modified 
blackbody spectrum and a value
  for $\beta$ \citep[e.g.][]{hatch07}, but given the assumptions involved we
  choose not to pursue this method.}  We will focus, therefore, on
$T_{bol}$ and $\alpha_{IR}$.

\subsection{Comparing Classification Methods}\label{classcompsec}

In Figure~\ref{tbolalpha} we compare classifications based on
$\alpha_{IR}$ and $T_{bol}$ for the candidate embedded protostars in
Perseus, Serpens, and Ophiuchus.  Thick symbols indicate sources
associated with 1.1~mm emission, while thin symbols denote sources with
upper limits at 1.1~mm.  The spectral index $\alpha_{IR}$ is
determined from a least squares fit to all detections between $\lambda=2\micron$ 
and  $\lambda=24~ \micron$.
The calculation
of $T_{bol}$ is described in \S\ref{bolsec} and \S\ref{appendsect}.
We adopt $T_{bol}$ class divisions from \citet{chen95} (see Table~\ref{deftab}), 
and $\alpha_{IR}$ divisions from \citet{am94}:
$\alpha_{IR} < 0$ (Class~I) and $\alpha_{IR} > 0$ (Class~II).  No
well-defined $\alpha_{IR}$ criteria exists for Class~0 sources, as
deeply embedded  objects were generally not visible in the mid-IR prior
to \textit{Spitzer}.

The two classification methods agree fairly well for Class~II and
``warmer'' Class~I sources.  Sources with $T_{bol} \lesssim 100$~K, by
contrast, have a wide range of $\alpha_{IR}$ values, and a few of
these ``cold'' sources even fall into Class~II based on the spectral
index.   One of the reasons for this large scatter in $\alpha_{IR}$
becomes apparent when examining the SEDs of some deeply embedded
sources,  which show considerable differences at short wavelengths.  Many
protostellar SEDs are not monotonically increasing from $3.6$ to $24~
\micron$, often falling from 5.8 to $8~ \micron$ and rising again at
longer wavelengths (e.g.,  Per-emb~18, Per-emb~22,  Oph-emb~1; 
Figure~\ref{psp3color}).  Geometric effects such as scattered light from 
an outflow cavity or absorption from ices in the protostellar envelope 
are the most likely cause of these features.  In any case,
non-monotonic behavior at short wavelengths will clearly bias the
calculation of $\alpha_{IR}$.  Calculating  $\alpha_{IR}$  from a
straight line fit between 2 and $24~\micron$, rather than a
least squares fit, still results in a large range of $\alpha_{IR}$
values at low $T_{bol}$.

To determine whether $\alpha_{IR}$ or $T_{bol}$ is a more accurate
measure of the true evolutionary state, we look at the correlation
between both measures and the degree to which a given source is
embedded.  Figure~\ref{distfig} shows the distribution of $T_{bol}$
with respect to the distance from each protostellar candidate to the
nearest 1.1~mm core, in units of the core FWHM size.  In all three
clouds, essentially all sources with $T_{bol} \lesssim 200$~K are
located within one core FWHM of a 1.1~mm peak.  Thus the majority of
``cold'' objects, as defined by low $T_{bol}$, appear to be embedded
in dense envelopes, and are likely to be at an early evolutionary
stage.  Furthermore, there is a clear correlation between smaller
distance and lower $T_{bol}$, suggesting that sources with lower
$T_{bol}$ are more embedded than those with higher $T_{bol}$.  This
correlation remains even if we exclude the 1.1~mm flux from the
calculation of $T_{bol}$, eliminating the possible bias between
millimeter cores and $T_{bol}$.  Thus, if we affiliate the degree of
embeddedness with youth, then $T_{bol}$ appears to correlate well with
evolutionary state.

Figure~\ref{adistfig} is similar, for $\alpha_{IR}$;
while there is some correlation between higher $\alpha_{IR}$ values
and smaller distances to the nearest 1.1~mm core, the relationship 
is not nearly as clear as for
$T_{bol}$.  In particular, there are a number of sources with large
$\alpha_{IR}$ values that do not seem to be embedded (the distance to
the nearest core is much larger than one FWHM), and there are several
deeply embedded sources with low $\alpha_{IR}$ values.
Based on Figures~\ref{tbolalpha}--\ref{adistfig} and  a visual
examination of sources with $T_{bol} < 70$~K, a number of
which are known Class~0 sources, we conclude that $T_{bol}$ is more
reliable than $\alpha_{IR}$ for classifying deeply embedded
protostars.  Hereafter,  $T_{bol}$ will be used to characterize
protostellar sources, and we use the divisions from \citet{chen95}
and Table~\ref{deftab} to place sources into Class~0, Class~I, or Class~II.

\section{Characteristics of Class 0, I, and II sources}\label{charsec}

\begin{figure*}[!ht]
\includegraphics[width=7.in]{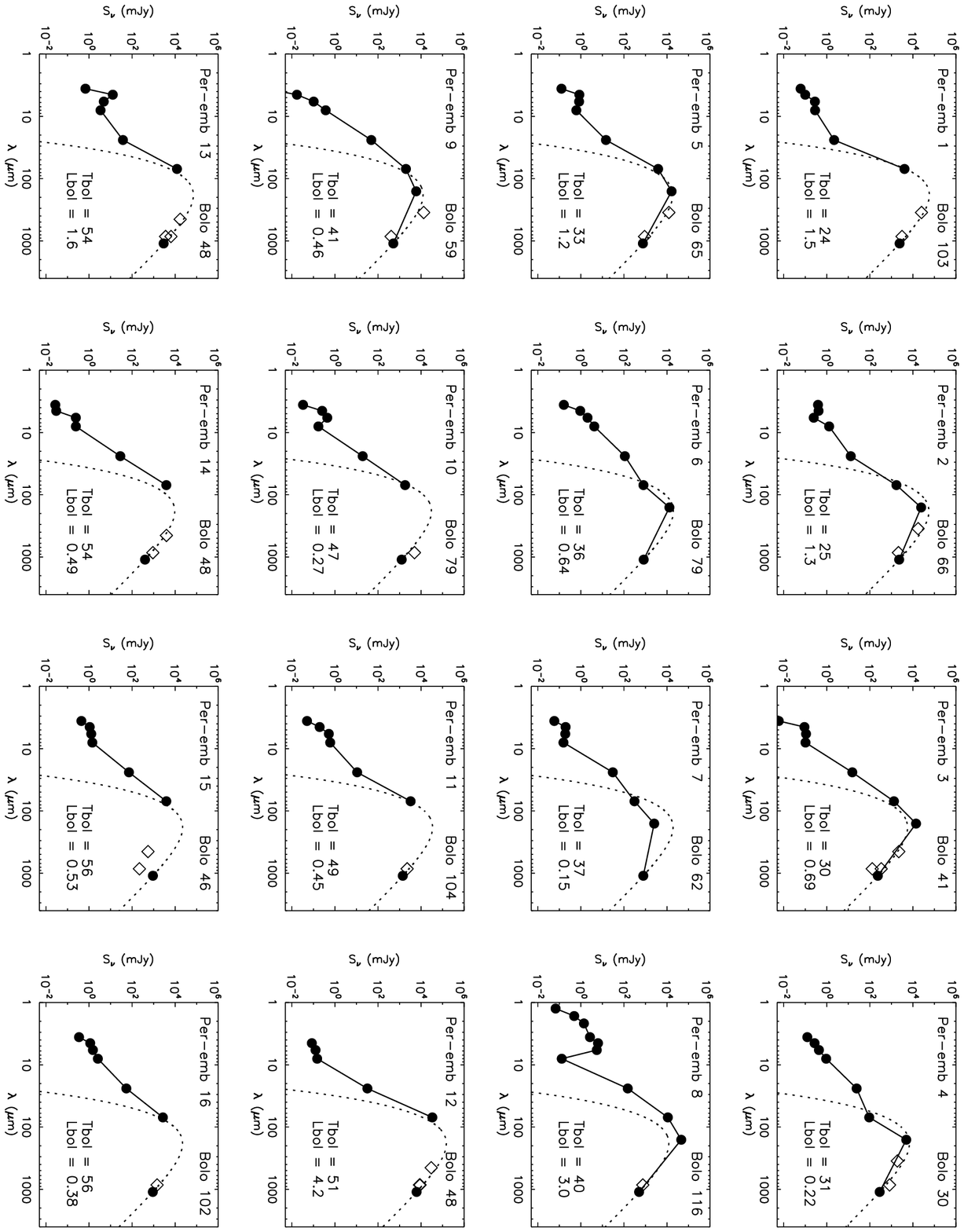}
\caption{Spectral energy distributions of all embedded protostars in 
Perseus, Serpens, and Ophiuchus, including photometry from 
2MASS (1.25, 1.65, 2.17~$\micron$), IRAC (3.6, 4.5, 5.8, 8.0~$\micron$), 
MIPS (24, 70, 160~$\micron$), and Bolocam (1.1~mm) data.  
Also included are SCUBA 450 and $850~\micron$ fluxes from the literature 
\citep{sk01,kirk06} and IRAS 25 or $60~\micron$ fluxes where they are used 
to replace saturated 24 or $70~\micron$ fluxes (see \S~\ref{bolsec}).
A sample is shown here; the full figure is available online only.
}\label{sedsfig}
\end{figure*}

The total numbers of Class 0 and Class I sources in each cloud, as
defined by $T_{bol}$ classifications, are given in
Table~\ref{lifetab}.  Note that in addition to $T_{bol}$, we also
require Class~I sources to be detected at 1.1~mm.  This is not an
unreasonable requirement, as our definition of Stage~I  dictates that
they should be detected at millimeter wavelengths ($M_{env} > 0.1
~\msun$).   The individual spectra (for $\lambda=1.25-1100 ~\micron$)
of all  Class 0 and Class I sources are shown in Figure~\ref{sedsfig}.

Numbers in parentheses in Table~\ref{lifetab} are based on
classifications using the prismoidal,  rather than midpoint, method
for calculating $T_{bol}$; these give some idea of  the uncertainties
in the number of Class~0 and Class~I sources.  Despite the sometimes
substantial difference in the midpoint and prismoidal $T_{bol}$ values
for individual sources, there is very little difference in the
resulting number of Class~0 and Class~I protostars, with the
exception of Serpens where the number of Class~0 sources increases
from 10 to 14.

Statistics for Class~II are not
given; as we intended to select against sources without a protostellar
envelope, we are necessarily incomplete to these objects.   For
example, there are 46 sources in our original Serpens candidate sample
with  $T_{bol}>650$~K, while \citet{harv07b} find 132 Class~II
sources.  We do detect a few Class~II sources at 1.1~mm, however (see
\S\ref{class2sec}).

\begin{figure*}[!ht]
\includegraphics[width=7.in]{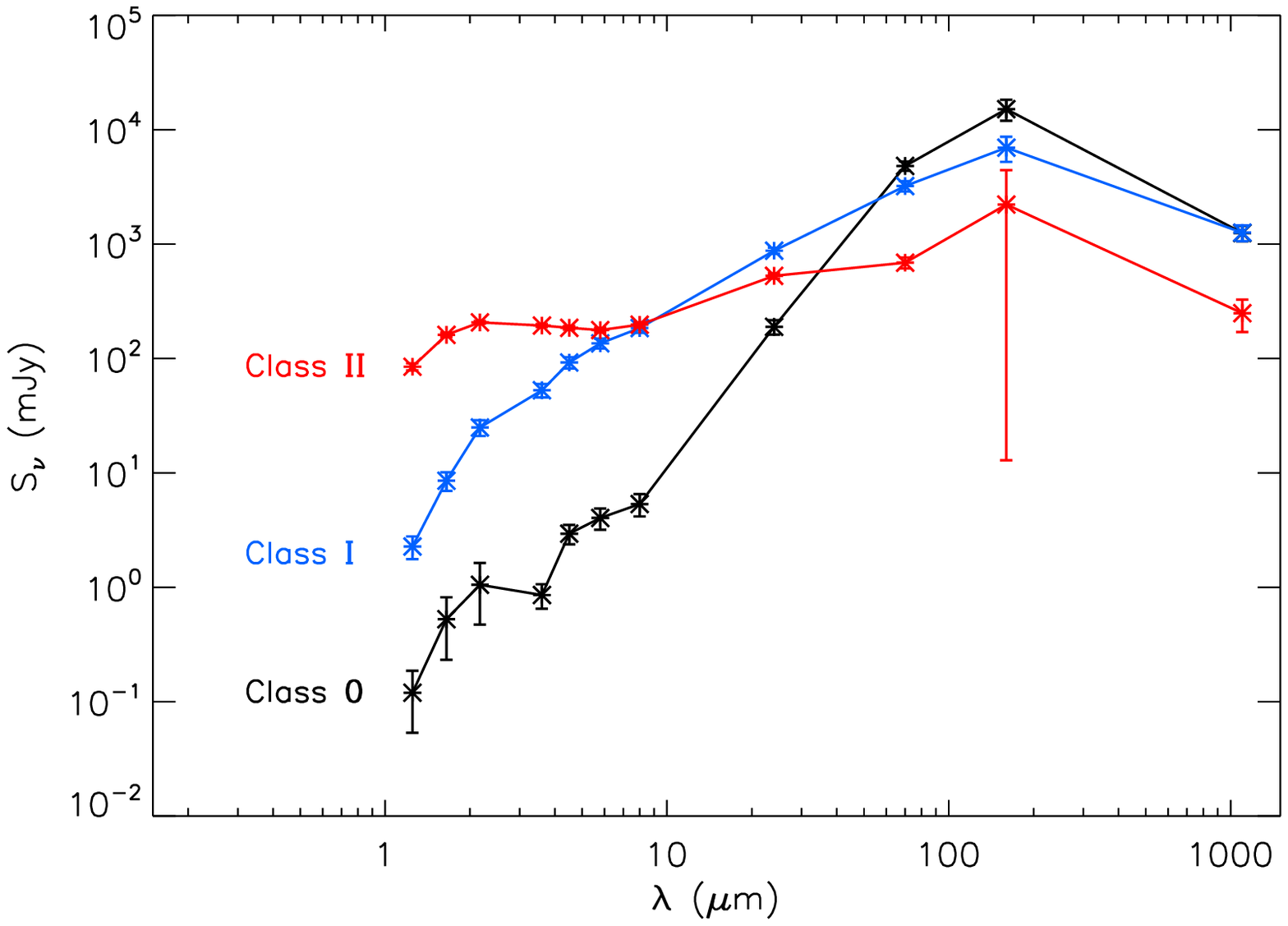} 
\caption
{Average spectra of sources in Class 0, Class I, and Class II,  for
  the Perseus and Serpens samples.   Sources are classified based on
  bolometric temperature, with the additional requirement that 
  Class~I sources be detected at 1.1~mm (see Table~\ref{deftab}).  
  Individual SEDs are weighted by $1/L_{bol}$ in the average
  calculation.   Error bars are the $1\sigma$ error in mean, and not
  the sample dispersion, which is much larger.   The Class~II spectrum
  is not necessarily representative of all Class~II objects,  as we
  are very incomplete to sources without protostellar envelopes.
}\label{avespec2}
\end{figure*}

The average spectra of Class~0, Class~I, and Class~II sources in
Perseus and Serpens are shown in Figure~\ref{avespec2}.  To calculate
the average spectra, individual SEDs are weighted by $1/\lbol$, so that 
we are not biased by the most luminous sources.  Error
bars in Figure~\ref{avespec2} represent the $1\sigma$ error in the
mean ($\sigma_{\lambda}/\sqrt{N_{\lambda}}$), but the dispersion in
the sample ($\sigma_{\lambda}$) is much larger.  Source SED shapes do
not fall into discrete bins, but form a continuous distribution
between the averages shown.   Error bars are large at wavelengths
where many sources are not detected (at short wavelengths for Class~0,
long wavelengths for Class~II, and $160\micron$ for all classes), and where
there are significant variations from source to source.

If Ophiuchus sources are included in the average calculation, the
resulting average spectra are skewed toward having more flux at short
wavelengths, have larger dispersion, and the similarity of SEDs within
each bin is reduced.  This behavior suggests that $T_{bol}$ may be
biased for many sources in Ophiuchus, likely due to the lack of
information at $160~ \micron$, so we exclude Ophiuchus
from the average spectra.

The progression from Class~0 to Class~I to Class~II is consistent with
a sequence of physical evolution.  In particular, the average Class~0
spectrum has the lowest flux densities from $1.25 - 24~ \micron$,
as expected for deeply embedded sources with massive, extincting
envelopes, and the highest fluxes at $70-1100~ \micron$, where
reprocessed protostellar flux is emitted by the cold envelope.  By
contrast, the average Class~II spectrum is relatively flat, with a
much larger percentage of the protostar flux emerging at shorter
wavelengths, as expected for older sources without much
circumstellar material.  One must keep in mind, however, that $T_{bol}$ 
is defined such that spectra which peak at longer wavelengths will have a lower $T_{bol}$.

Large error bars on the average Class~0
spectrum for $\lambda=1-3~ \micron$ are indicative of the widely
varying behavior of Class~0 objects in the near-infrared, and the
presence of non-monotonic behavior is apparent  at $\lambda = 3.6~
\micron$.  
The large separation between the Class~0
and Class~I spectra, as well as the continuous range of SED shapes,
suggests that a more continuous means of estimating evolutionary
status is preferable to the standard classes.

\subsection{Class II Sources with 1~mm Emission}\label{class2sec}

The average Class~II spectrum
is not necessarily representative of all Class~II objects,  due to
severe incompleteness to non-embedded sources.
The Class~II sources we do detect are likely relatively young Stage~II objects, 
before a substantial fraction of disk mass is dispersed or accreted. 

Given our 1.1~mm sensitivity limit of approximately $0.1~ \msun$,
almost all sources detected at 1.1~mm will be dense cores or envelopes
around relatively young protostars.  In general, by the time a
protostar has consumed or dispersed its massive envelope  and enters
Stage~II (or the T~Tauri phase, e.g., \citealt{als87}), the remnant 
disk of gas and dust has too little mass to be detected by our millimeter 
surveys.  \citet{crapsi08} define Stage~II as sources with a 
circumstellar mass below $0.1~\msun$.  Typical measured masses of
Class~II disks are $0.01-0.1~ \msun$ (e.g., \citealt{bs96}), although
values as large as $1~ \msun$ have been measured \citep{beck90}.

In a few a cases we do detect 1.1~mm emission around sources
with Class~II-type SEDs.  These objects have bolometric temperatures
$T_{bol} > 650$~K and a
flux density at 1.1~mm that is lower than the flux densities from
$3.6$ to $24~ \micron$.   With the exception of two ``flat spectrum''
objects ($-0.3<\alpha_{IR}<0.3$; \citealt{greene94}), the near- to
mid-infrared spectral indices, $\alpha_{IR}$, of these objects are in
the range $-0.34$ to $-1.04$,  confirming their Class~II status.  In
some cases, the 1.1~mm emission is unresolved, consistent with a
compact disk.  Often, however, these Class~II sources are in a region
of confused millimeter emission, so their physical association with
the 1.1~mm emission is not secure.  

Table~\ref{cl2tab} lists all
Class~II sources ($650<T_{bol} <2800$~K) in each cloud that are detected 
in our 1.1~mm surveys.  Those in confused regions of 1.1~mm emission
are indicated by a ``$^*$'' in the ``Bolocam ID'' column. 
Sources with point-like 1.1~mm emission 
emission centered on the \textit{Spitzer} position
(3 in Perseus, 1 in Serpens, and 3 in Ophiuchus) 
may have massive disks.

\subsection{Individual Sources}\label{indsec}
Here we briefly discuss a few examples of newly-identified or otherwise 
interesting embedded protostellar sources.

\subsubsection{IRAS 03292+3039}

\begin{figure*}[!ht]
\vspace{0.6in}
\begin{center}
\includegraphics[width=6.in]{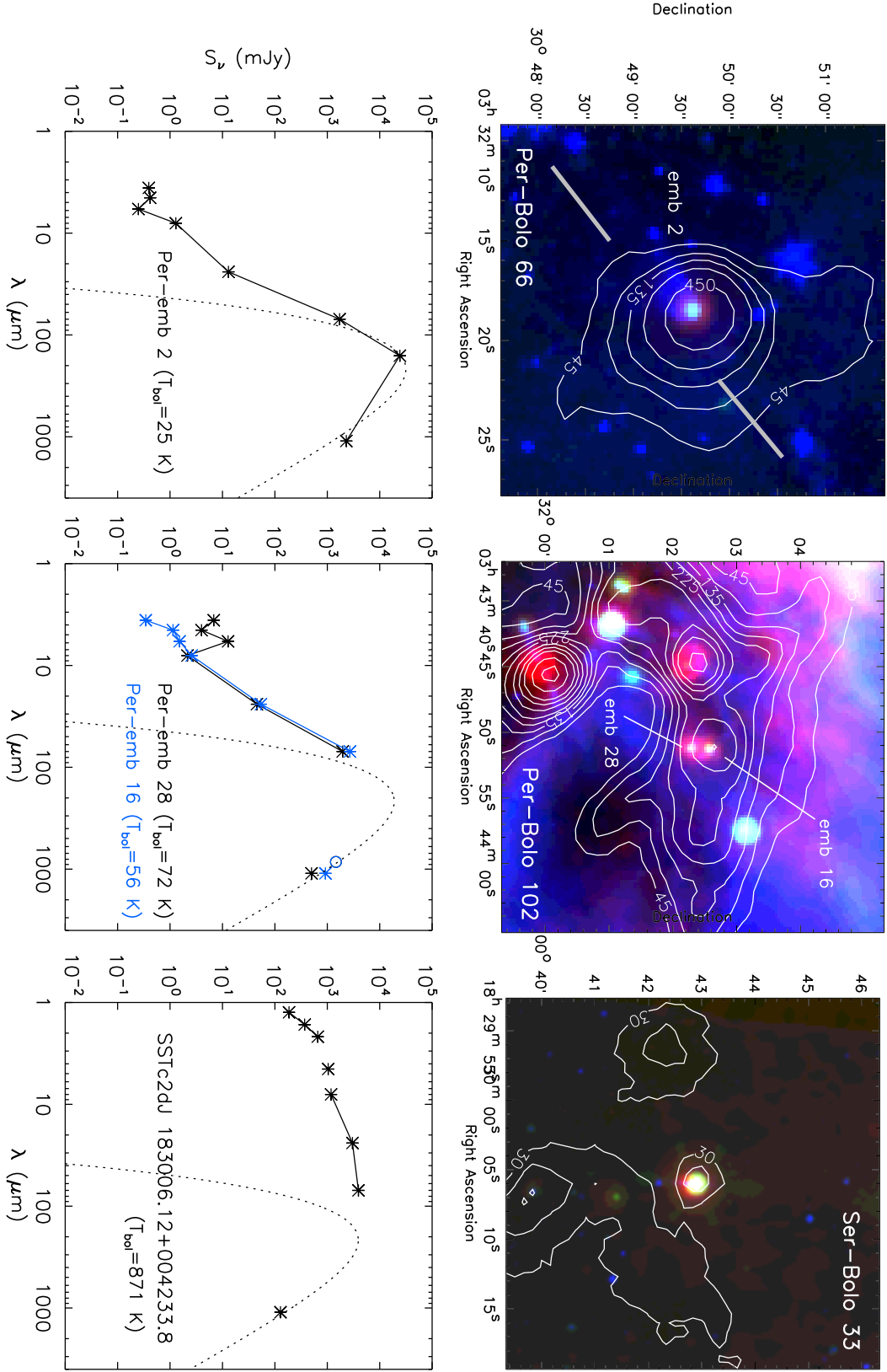}
\caption
{Three-color \textit{Spitzer} images and SEDs of individual sources
  discussed in \S\ref{indsec}.  Images are ($8,24,70~ \micron$), unless
  otherwise noted, with 1.1~mm contours. 
  Left: Three-color ($3.6,24,70~ \micron$)
  image of IRAS 03292+3039 (Per-emb~2).  Fan-shaped nebulosity is
  visible at $3.6~ \micron$,  most likely scattered light from a
  narrow outflow cavity in this deeply embedded Class~0 source.  Thick
  lines indicate the approximate orientation of the larger-scale
  outflow \citep{jorg06}.  Center:  Per-Bolo~102, a new
  candidate binary Class~0/I source.  Although the SEDs of the two
  embedded protostars look very similar at $\lambda \ge 8~ \micron$,
  the southern source is brighter at shorter wavelengths, fainter at
  2.7~mm (see text), and has a slightly higher $T_{bol}$.
  Right:  Ser-Bolo~33, a bright Class~II object (Table~\ref{cl2tab}).
  While this is clearly a more evolved source, it is massive enough that we detect
  1.1~mm emission from a compact disk.}
\label{indfig}
\end{center}
\end{figure*}

IRAS~03292+3039 (Per-emb~2) is a little-studied, Class~0 source
associated with the 1.1~mm core Per-Bolo~66; it was discussed
briefly in Paper~I.  An image of IRAS~03292+3039, together
with the SED, is shown in Figure~\ref{indfig}.  \citet{jorg06}
identified this as an outflow source, noting the large-scale
outflow visible in the $4.5~ \micron$ IRAC band.

Bright 1.1~mm emission centered on this object indicates a fairly
massive protostellar envelope (2.9~\msun), while the powerful outflow
\citep{wal05,jorg06,hatch07b} and low bolometric temperature ($T_{bol} = 25$~K)
are evidence of an extremely young, energetic embedded protostar.  The
spectral energy distribution  is similar to well-known Class~0
protostars in Perseus such as NGC~1333-IRAS~4, and nearby 
IRAS~03282+3035.  The small fan-shaped nebulosity visible at 3.6 and
$4.5~ \micron$ (Figure~\ref{indfig}) is most easily explained by a
cone-shaped cavity, carved out of the dense envelope by an energetic
outflow.  The orientation of the nebulosity corresponds well to the
larger-scale outflow traced by IRAC emission \citep{jorg06}, which is
indicated by thick gray lines in Figure~\ref{indfig}.  The one-sided
nebulosity at $3.6~ \micron$ is strikingly similar to the
\citet{whit03} model IRAC image of an early Stage~0 sources viewed at
an inclination angle of  $30^{\mathrm{o}}$ (their Figure~12a).

\subsubsection{Per-Bolo 102}

Per-Bolo~102 is a bright 1.1~mm source that was identified in the
Bolocam survey of Perseus \citep{enoch06}.  It lies within the region
of active star-formation near IC~348, which includes the famous
outflow-driving source HH~211 \citep{mcg94}.  The \textit{Spitzer}
$24~ \micron$ image resolves the luminous internal source into a
double object (Per-emb~16 and Per-emb~28).  While the two SEDs are
very similar for $\lambda \ge 8 ~ \micron$ (Figure~\ref{indfig}),
variations at shorter wavelengths cause Per-emb~16 ($T_{bol} =56$~K)
to fall into Class~0, while Per-emb~28 ($T_{bol} =72$~K) falls just
outside the Class~0/Class~I boundary. These
sources are good examples of why a more continuous evolutionary scheme
is preferable to the standard classifications.  

Recently obtained CARMA interferometric observations at
$\lambda=2.7$~mm resolve the millimeter core into two sources,
coincident with the \textit{Spitzer} source positions (M. L. Enoch et al.,
in preparation).  The  flux ratio at 2.7~mm of the northern (Per-emb~16) to
southern  (Per-emb~28) source is at least 2:1, further evidence that
Per-emb~28 is  slightly more evolved.  Although not as massive or cold
as nearby HH~211 (Per-emb~1) and IC~348-mms (Per-emb~11), Per-Bolo~102
is an interesting case study.  It may be a separate-envelope binary
system, with two nearly coeval Class~0 or early Class~I sources.   The
sources are  separated by 17\arcsec, or 4200~AU.  Binary separations
of this order are consistent with early fragmentation in a relatively
dense cloud (``prompt initial fragmentation'', e.g.,
\citealt{pring89,loon00}), in which case the individual sources would
have distinct protostellar envelopes.  

In a binary formed via gravitational fragmentation, we would expect
the separation to correspond to the local Jeans length
\citep{jeans28}:
\begin{equation}
\lambda_J = \left( \frac{\pi c_s^2}{G \mu_p m_{H} n} \right)^{1/2},
\end{equation}
where $c_s$ is the local sound speed, and $\mu_p=2.33$ and $n$ are the
mean molecular weight and mean particle density, respectively.  A
Jeans length of 4200~AU would require a relatively high density ($n
\sim 6\times10^5$~cm$^{-3}$, assuming $c_s=0.2$~km~s$^{-1}$).  The
mean density of the Per-Bolo~102 core, measured within an aperture of 
$10^4$~AU, is $4\times 10^5$~cm$^{-3}$, close to the required value.

\subsubsection{Ser-Bolo 33}

One noteworthy example of a Class~II source associated with 1~mm 
emission is Ser-Bolo~33 (SSTc2dJ183006.12+004233.8)
in Serpens (Figure~\ref{indfig}), a very bright
\textit{Spitzer} source with $L_{bol}=3.6~ \lsun$, $T_{bol} = 871$~K,
and compact 1.1~mm emission centered on the \textit{Spitzer} position.
The spectral index, $\alpha_{IR} = -0.42$, also places this object in 
Class~II.  \citet{vie03} included this source in a sample of
Herbig Ae/Be candidates,  believed to be the intermediate mass
Class~II counterparts of low-mass  T~Tauri objects, although the
spectral type is F3.  The 1.1~mm mass calculated assuming an optically
thin disk is $0.17~ \msun$, approximately 10\% of the stellar mass,
$M_{*} \sim 1.3 ~ \msun$, estimated from the measured effective
temperature ($T_{\mathrm{eff}} \sim 6300$~K,  \citealt{vie03}) and an
empirical $T_{\mathrm{eff}}-M$ relation \citep{hh81}.

\begin{figure*}[!ht]
\includegraphics[width=7.3in]{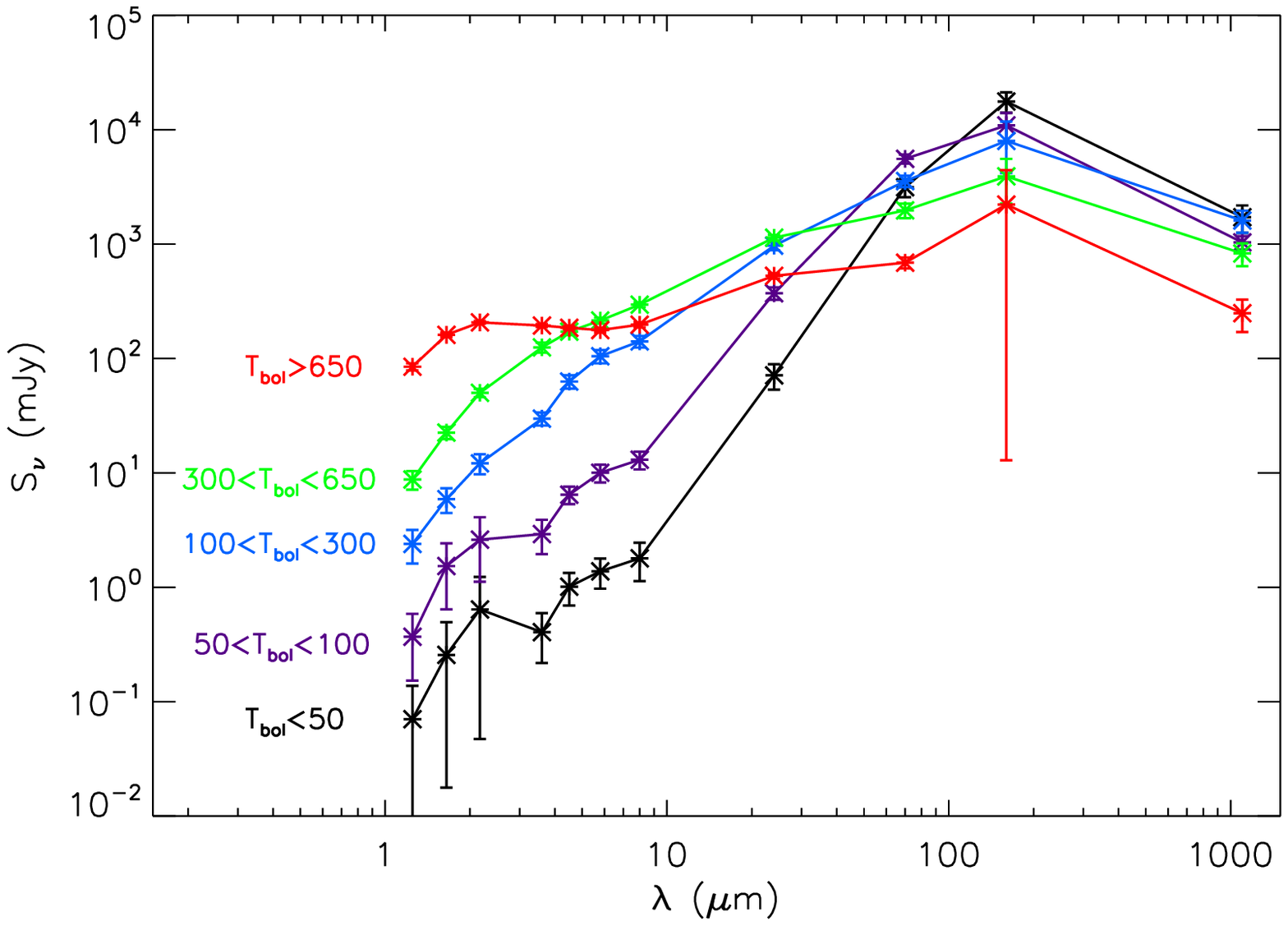} 
\caption
{Average spectra in designated $T_{bol}$ bins (see Table~\ref{deftab}), 
using sources from
  Perseus and Serpens.  The average is calculated as described in
  Figure~\ref{avespec2} and the text, and error bars represent the
  error in the mean.  The progression in SED shape for increasing
  $T_{bol}$ is as expected if this represents physical evolutionary
  sequence.  
}\label{avespec}
\end{figure*}

\section{Alternative Classifications}\label{otherclasssec}

One advantage of such a large sample is that it allows us to
define a more  continuous evolutionary sequence than the standard
classes that were appropriate for the smaller samples previously
available.  With this in mind, we divide the protostellar sources in
each cloud into smaller $T_{bol}$ bins than those of the standard
Class~0/I/II divisions.  
Average spectra for ``early Class~0'', ``late Class~0'',  ``early Class~I'', 
``late Class~I'', and Class~II (see Table~\ref{deftab} for definitions) 
sources in Perseus and Serpens
are shown in Figure~\ref{avespec}.  Error
bars represent  the $1 \sigma$ error in the mean, and average spectra
are calculated as described in the previous section.  The largest
error bars are seen for the shortest wavelengths in the lowest $T_{bol}$
bins, where NIR fluxes vary significantly from source to source,
likely depending on outflow opening angle and viewing geometry.
Binning of $T_{bol}$ is based on general agreement of SEDs in a given
bin, as determined by eye, and should not be interpreted as strict
boundaries.

While there are significant similarities between the SEDs within each
bin, individual source SEDS fill the continuum between the average
spectra, and each average
spectra lies within the $1\sigma$ dispersion of neighboring $T_{bol}$ bins.
The average spectra change as one would expect  if an extincting
envelope is gradually accreted or dispersed, with  the protostar
becoming more visible at short wavelengths.   At all wavelengths
except 24 and $70~ \micron$, the flux rises or falls monotonically
with increasing $T_{bol}$.  For these intermediate wavelengths, the
observed flux may rise initially as hotter dust close to the protostar
is revealed, then fall as the mass of circumstellar material drops.
Despite the relatively narrow bins, there is still a rather large
change between ``late Class~0'' and ``early
Class~I'', particularly at $\lambda=3.6-24~\micron$;
this transition may occur rapidly, or these wavelengths may be 
particularly sensitive to geometry.

\section{Comparison to Models}\label{modsec}

Even for infinitely
well-sampled SEDs, spectrum shape is not necessarily directly
correlated with age or degree of embeddedness. For example, viewing
geometry can have a strong effect on SED shape.  In three
dimensional radiative transfer models of \citep{whit03,rob06,crapsi08}, 
Stage~I and Stage~II sources can have quite similar SEDs when viewed at the right
inclination angle (e.g., when the observer's line-of-sight intersects
the outflow opening angle of Stage~I sources).  Even Stage~0 sources
can appear much warmer if we happen to be looking directly into the
outflow cavity, although the probability of that occurring is small.

\begin{figure*}
\begin{center}
\includegraphics[width=6.5in]{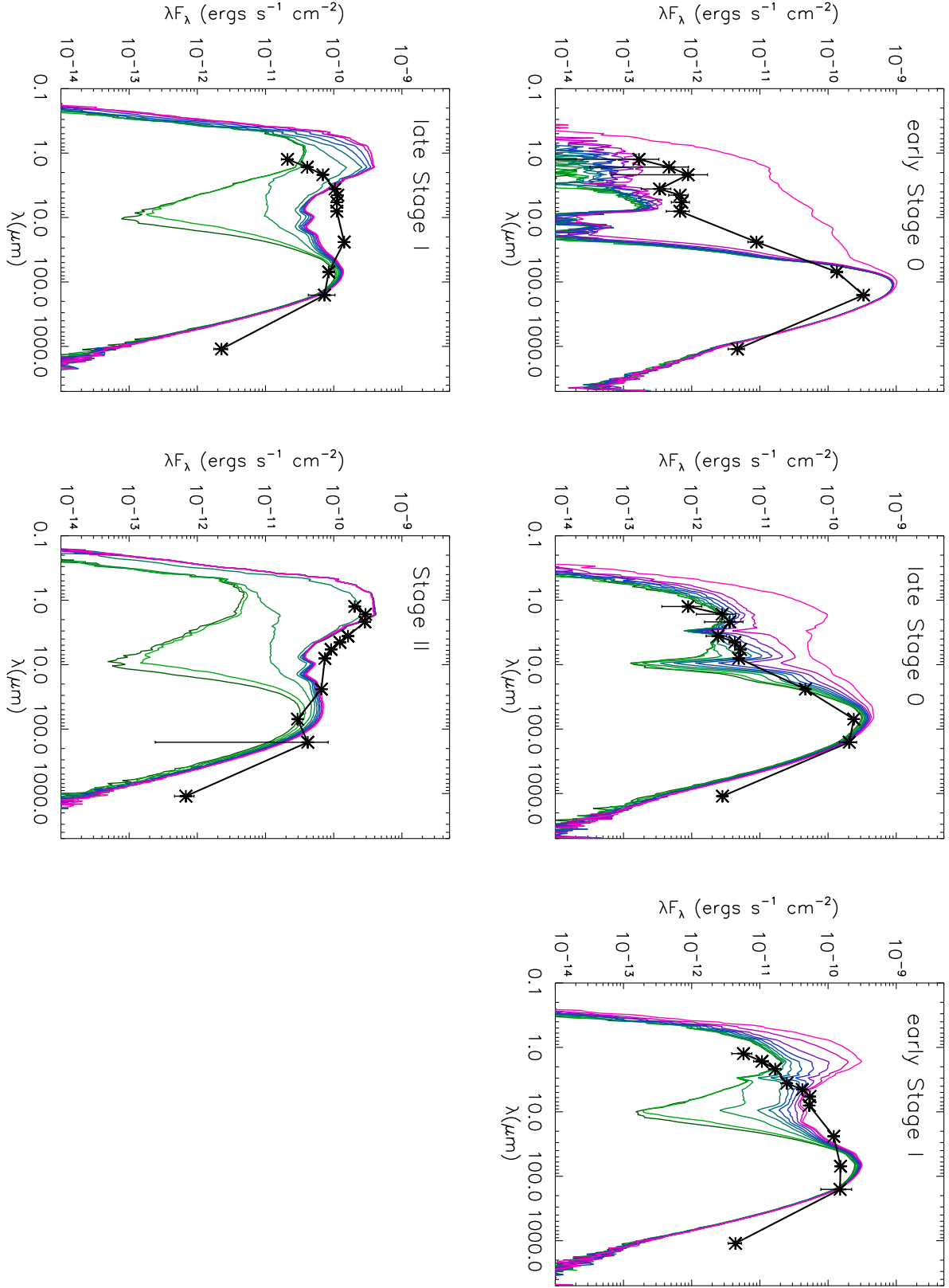}
\caption
{Comparison of the average observed spectra from Figure~\ref{avespec}
  with predicted spectra from the protostellar evolution models of
  \citet{whit03} (plotted as $\nu S_{\nu}$).  Distinct geometries are chosen to represent five
  evolutionary stages, and include a combination of an accreting
  protostar, a protostellar disk, envelope, and a bipolar outflow.
  Colors indicate inclinations angles of the protostellar disk, from
  $0^{\mathrm{o}}$ (pink) to $90^{\mathrm{o}}$ (dark green).  With some exceptions the
  Stage~0 and Stage~II models match the observed spectra fairly well,
  but the Stage~I models over-predict the near-IR and under-predict
  the mid-IR flux.  In general the models tend to underestimate the
  millimeter flux and thus the mass of the protostellar envelope.
}\label{modfig}
\end{center}
\end{figure*}

We compare our spectra to the results of protostellar models, which
predict protostellar spectra based on source age, mass, accretion
rate, etc., both to gain insight into the evolutionary state of
sources and to evaluate the effectiveness of such models in matching
observed sources.  Rather than model each source individually, we
compare the average spectra from Figure~\ref{avespec} with predicted
spectra from \citet{whit03} for ``early Stage 0'', ``late Stage 0''
sources, etc. in Figure~\ref{modfig}.  

\citet{whit03} begin with a set
geometry for each stage, including some combination of accreting
protostar, flared protostellar disk, infalling envelope, bipolar outflow, and grain
models for each region, then use radiative transfer modeling to
predict protostellar spectra.   Colors in Figure~\ref{modfig}
correspond to inclination angles of the protostellar disk from $0^{\mathrm{o}}$
(pink) to $90^{\mathrm{o}}$ (green), where the outflow is perpendicular to the
disk.  Model envelope infall rates decline from $10^{-4}$ \msun
yr$^{-1}$ in early Stage~0 to $10^{-6}$ \msun yr$^{-1}$ in late Stage~I.
Similarly, the disk radius and cavity opening angle increase and the 
cavity density decreases as one moves from early Stage~0 to late Stage~I.
Both models and average observed spectra (thick black lines) are scaled 
to a total luminosity of $1 \lsun$.

The model early Stage~0 and observed early Class~0 
spectra agree fairly well.  The observed spectrum lies above most of
the models at $\lambda \le 24 \micron$, which could be explained by
gaps in the inner envelope of some protostars, allowing short
wavelength flux to escape \citep{jorg05}, or a wider outflow angle
than the model ($5^{\mathrm{o}}$).  The \citet{whit03} models predict that a
small fraction (approximately $1/10$) of early Stage~0 sources will
have spectra similar to Stage~I, at very low inclination angles
(looking down the outflow).  The observed Class~II spectrum is also
consistent with at least the low inclination models; as we select for
sources with 1.1~mm emission it is not unexpected that the observed
millimeter point is higher than predicted by the models. 
For late Stage~0 the agreement is again pretty good, except that the
model under-predicts the millimeter flux, and thus the mass of the 
protostellar envelope, a feature present in all
spectra later than early Stage~0.

\begin{figure*}[!ht]
\vspace{-0.6in}
\hspace{-0.3in}
\includegraphics[width=7.2in]{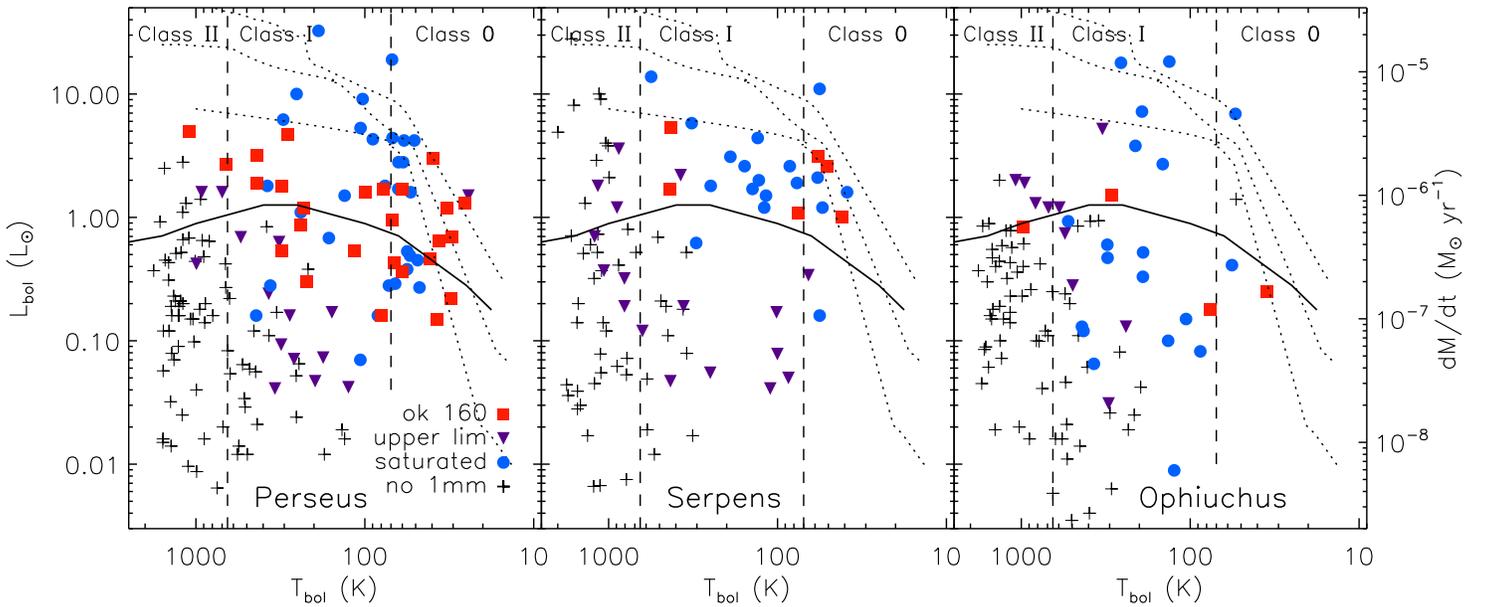} 
\vspace{-0.3in}
\caption
{Bolometric luminosity versus bolometric temperature
  ($L_{bol}-T_{bol}$) diagram for embedded protostar candidates in
  Perseus, Serpens, and Ophiuchus. Filled and open symbols indicate
  that a given source is associated with 1.1~mm emission, while ``+''
  symbols indicate upper limits at 1.1~mm.   Symbol are further
  divided based on whether there is a reliable $160 \micron$
  measurement,  an upper limit, or if we are unable to measure a $160
  \micron$ flux due to saturation.  
  The right axis shows the mass accretion rate, $dM/dt$, calculated 
  from $L_{bol}$ using Eq.~\ref{dmdteq} and assuming $M_* = 0.5 M_{\odot}$ 
  and $R_*=5R_{\odot}$.  Model evolutionary tracks from
  \citet{ye05} for sources with 3.0, 1.0, 0.3~$M_{\sun}$ (dotted
  lines,  from top to bottom) and from \citet{myer98} for a source
  with final mass 0.3~$M_{\sun}$ (solid line) are shown for
  comparison.  
  The large population of low-luminosity ($L_{bol} < 0.1 \lsun$) Class~I 
  objects argues strongly for episodic accretion by Stage~I. 
}\label{bols1}
\end{figure*}

By contrast, the agreement between
both Stage~I models and the observed average spectra is quite poor.
The models severely over-predict the near-IR flux and under-predict
the mid-IR flux.  In fact, the characteristic double-peaked profile of
the Stage~I models is almost never seen in the observed spectra (Figure~\ref{sedsfig}). 

\citet{ind06} demonstrate that envelopes with clumpy density
distributions can increase the observed flux at $10~\micron$, with
short wavelength emission escaping through low opacity regions between
the clumps. This kind of clumpy distribution eliminates some of the
``double-peaked'' structure apparent in the \citet{whit03} models,
which assume a smooth density profile in the envelope and constant
density in the outflow cavity, although the clumpy models also
increase the flux at $<3\micron$.
Alternatively, the presence of a larger, thicker 
protostellar disk could flatten the Stage~I spectra by absorbing near-IR 
flux and producing more mid-IR flux.  A foreground cloud could also 
absorb near-IR flux, without producing any mid-IR flux. 

It seems that we still do not understand how sources transition
from Stage~0 to Stage~I, at least in relation to these models.  
Whether it is the physical model that is unrealistic or how that 
physical geometry translates to the observed spectra is not clear. 
It is important to note that the same
authors have developed a more complete grid of models \citep{rob06};
we compare here to the earlier models rather than fitting
each average spectrum because we want an idea of the
global agreement with evolutionary stage.

\section{Protostellar Evolution}\label{evolsec}

\subsection{Luminosity Evolution}\label{levolsec}

We examine the evolution of embedded protostars in more detail using the
$L_{bol}-T_{bol}$ diagram, the protostellar equivalent of the
H-R diagram \citep{myer98}.  Here $T_{bol}$ is used as a measure of
temperature rather than $T_{eff}$, which is not well-characterized for
embedded sources.  As discussed in \citet{myer98}, newly-formed
protostars should begin at low $L_{bol}$ and $T_{bol}$, increasing in
both $L_{bol}$ and $T_{bol}$ as accretion proceeds.  If accretion
abates or is otherwise halted, then $L_{bol}$ will decrease for
steadily increasing $T_{bol}$.  Eventually, sources will move onto the
main sequence, at $T_{bol}\gtrsim3000$~K.

Figure~\ref{bols1} plots $L_{bol}$ versus $T_{bol}$ for the candidate
embedded protostar samples in Perseus, Serpens, and Ophiuchus.
Filled and open symbols are used for sources that are associated with
1.1~mm emission, while ``+'' symbols indicate sources with upper
limits at 1.1~mm.\footnote{Note that although we plot sources with
  1.1~mm upper limits here, our final embedded protostar  samples
  (Tables~\ref{boltabp}--\ref{boltabo}) includes only sources with
  1.1~mm detections.}     All sources detected at 1.1~mm are further
divided according to whether they have reliable measured $160~
\micron$ fluxes (squares), are  saturated at $160~ \micron$
(circles), or are not detected at $160~ \micron$ (triangles).
Given the discussion in \S\ref{appendsect} regarding the effect of
missing $160~ \micron$ fluxes on the calculation of $T_{bol}$, circles
would be expected to move up and to the right in this diagram for
``cold'' sources ($T_{bol} \lesssim 100$~K), and down and to the left
for ``warm'' sources ($T_{bol} \gtrsim 100$~K), if $160~ \micron$
fluxes were available.  

While upper limits at 1.1~mm (``+'' symbols) are expected for more
evolved sources with $T_{bol}\gtrsim 500-600$~K, colder sources with
no 1.1~mm detection may be either misclassified (e.g., background
galaxies), or very low mass sources whose  1.1~mm flux is below our
detection limit.  Given our physical definition of Stage~I ($M_*\gtrsim
M_{env}$; $M_{env}>0.1 M_{\sun}$), embedded sources not detected at 1.1~mm
must have stellar masses less than a few tenths of a solar mass,  or have very
little remaining envelope ($M_* \gg M_{env}$).  

Protostellar evolutionary tracks, which predict source properties as a
function of age and mass, can easily be compared to our data using the
$L_{bol}-T_{bol}$ diagram.  Model tracks from \citet{myer98} (solid
line; hereafter M98) and \citet{ye05} (dotted lines; hereafter YE05)
are shown in Figure~\ref{bols1}.   YE05 adopt the standard inside-out
collapse model of \citet{shu77}, and assume that no mass is lost in
the formation process. From top to bottom, the YE05 models are for
sources with masses of 3.0, 1.0, and 0.3~\msun, assuming a constant
accretion rate of $dM/dt = c_s^3/G$, where $c_s$ is the effective
sound speed \citep{shu77}.  From the initial singular isothermal
sphere, finite  masses are achieved by truncating  the outer radius of
the envelope.  A one dimensional radiative transfer model (DUSTY) is
used to calculate observational signatures ($L_{bol}$, $T_{bol}$,
etc.)  from the accretion model.  

Unlike YE05, M98 do not assume that that the entire mass of the
original core ends up in the final star, but rather that a significant
fraction of the  core mass is lost in the star formation process.
The M98 model shown is for a source with initial core mass of $1.8~
\msun$ and final stellar mass of 0.3~$M_{\sun}$.  M98 assume an
accretion rate that is initially $dM/dt = c_s^3/G$, but falls off
exponentially with time, designed to match the observed luminosity of
pre-main sequence stars at $T_{bol} \gtrsim 3000$~K.  Thus, the
luminosity is significantly lower than the YE05 tracks at later times.
Both evolutionary models assume an accreting central protostar, a
circumstellar accretion disk, an extended envelope, and a contribution
to the luminosity from gravitational contraction of the protostar.
YE05 also include nuclear (Deuterium) burning.

Ignoring for a moment the population of Class~I sources in each
cloud with $L_{bol}$ values well below both models, the M98 model, for
which a large fraction of the core mass is ejected or dispersed, is
more consistent with the observed protostellar sources.  
The ratio of the final stellar mass to initial core mass for the M98
model shown ($f_{\mathrm{eff}}=M_{core}/M_{star} = 0.3/1.8 = 0.17$) is 
smaller than  the values ($f_{\mathrm{eff}}=0.3\pm0.1$) found by
\citet{alves07} and \citet{enoch08} by comparing the shape of the core 
mass distribution to the initial mass function ($f_{\mathrm{eff}}=0.3\pm0.1$ 
and $f_{\mathrm{eff}} \gtrsim 0.25$, respectively).

The slightly better match to the M98 model may be irrelevant, however, 
as neither a constant nor a
steadily decreasing accretion rate is consistent with the
observed protostellar populations in these clouds; many sources lie
below all four model tracks.  Our data confirm and exacerbate the
``luminosity problem'' noted by \citet{ken90}, that Class~I
protostars in Taurus were observed to have lower $L_{bol}$ values than
expected based on the average mass accretion rate required to make a
$1 \msun$ star.

In particular, the large population of  Class~I sources with low
$L_{bol}$ in each cloud is difficult to understand in relation to most
existing protostellar evolutionary models.  A general feature of such
models is that the bolometric luminosity peaks in the Class~I stage, a
result that is true for constant accretion rates (YE05), decreasing
rates (M98), and gravo-turbulent models \citep{fro06}.   In contrast,
we find quite similar mean luminosities for the Class~0 and  Class~I
samples:  2.4~\lsun and 2.2~\lsun, respectively.  The standard
deviation  for both samples is large: 3.5~\lsun\ and 4.7~\lsun\ for
Class~0 and  Class~I, respectively.  The median luminosity of Class~I
sources is a factor of 3 lower than  the mean, 0.7~\lsun.

\begin{figure*}[!ht]
\begin{center}
\includegraphics[width=6.in]{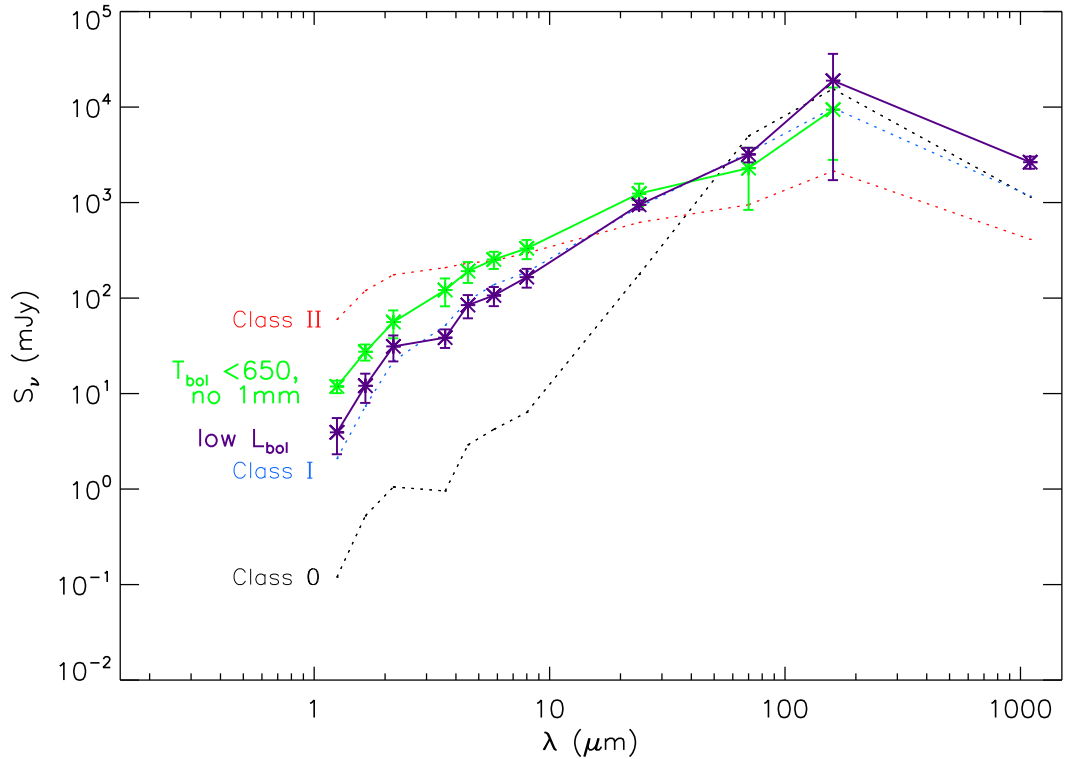}
\caption
{Average spectra of low-luminosity Class~I sources
  and ``non-envelope Class~I'' sources (those that have
  $T_{bol}<650$~K but no 1.1~mm emission).  Average Class 0, I, and II 
 spectra from Figure~\ref{avespec2} are shown (dashed lines) for reference.
  The low-$L_{bol}$ Class~I sources have lower short wavelength points
  relative to the  $160~\micron$ and 1.1~mm fluxes, as expected for
  low accretion rates,  but are otherwise consistent with the average
  Class~I spectrum.  The average ``non-envelope
  Class~I'' spectrum is intermediate between the average Class~I and
  Class~II spectra, as expected if these sources represent an
  transitional stage in which there is little or no remaining
  envelope.}
\label{avespec3}
\end{center}
\end{figure*}

\subsection{Episodic Accretion}\label{episodicsec}

Given the above discussion and the large observed spread in $L_{bol}$
of three orders of magnitude for Class~I sources, much larger than
the range in envelope masses, we conclude that mass accretion  during
the Class~I stage is episodic.  Class~I sources with low $L_{bol}$ can
be explained by periods of relative quiescence when the bolometric
luminosity, which is driven primarily by accretion luminosity, drops
by at least a factor of 10.  Conversely, Class~I sources with high
$L_{bol}$ values, e.g., those that form the upper envelope of the
distribution and appear to be consistent with the YE05 models, would
correspond to periods where the accretion rate is close to the ``Shu
accretion'' value.  Approximately 20\% of the Class~I sources have
$L_{bol} < 0.1 \lsun$ ($7/39$ in Perseus, $5/25$ in Serpens, and
$5/25$ in Ophiuchus), and could be considered candidate very low
luminosity objects (VeLLOs; \citealt{dunham08,dif07}).

Episodic accretion is not an unreasonable solution; evidence for
variable mass accretion and ejection is plentiful, including that
based on modeling FU Orionis eruptions \citep[e.g.,][]{hk85},   and
bow shocks in Herbig-Haro outflows \citep[e.g.,][]{rb01}.  
In traditional models of episodic accretion, infall from the envelope 
onto the disk is constant and accretion from the disk onto the protostar is episodic
\citep[e.g.,][]{kh95}, for example due to gravitational instabilities in the disk \citep{vb06}.
Other scenarios such as the ``spasmodic''
infall model of \citet{tm05}, where material is held up by the
magnetic field at the inner edge of the envelope are also plausible.

Although the observed distribution of $L_{bol}$ for Class~I sources
argues strongly for episodic accretion, that is not the only
possibility.   One alternative is that the  low-$L_{bol}$ Class~I
sources are simply very low mass objects, and we are somehow missing
their low-mass Class~0 counterparts.   Based on our sensitivity limits
at $70,160~ \micron$ and 1.1~mm, we should be able  to detect Class~0
sources with $L_{bol} \gtrsim 0.05\lsun$.  The lower limit to the
observed Class~0 luminosity, however, may not be the internal
luminosity, but heating of the envelope by the interstellar radiation
field (ISRF), which can contribute as much as $0.2-0.3 \lsun$ to
$L_{bol}$ for envelope masses of $1-3 \msun$ (YE05).  For the same reason, a
Class~0 source with a very low mass accretion rate would not
necessarily have a very low bolometric luminosity, making it difficult
to determine if episodic accretion is already present at the Class~0
stage.

In Figure~\ref{avespec3}, we
show the average spectrum of  low-luminosity ($L_{bol} < 0.2 \lsun$) 
Class~I sources in Perseus
and Serpens.  On average, these sources have suppressed short
wavelength points relative the  $160~\micron$ and 1.1~mm fluxes,  but
are otherwise  consistent in shape with the average Class~I
spectrum.\footnote{Note that the average spectra are normalized by
  $1/\lbol$, so on an absolute scale the low-\lbol\ spectrum would be
  substantially fainter than the Class~I spectrum.}
The average low-\lbol\ spectrum is consistent with low accretion
luminosities (evident at short wavelengths)  in sources with
``normal'' envelope masses, as expected if the low luminosity objects
are similar to other Class I sources  but with lower mass accretion
rates.  
The characteristics of low luminosity protostars in Perseus, Serpens, and Ophiuchus 
are analyzed in \citet{dunham08}.

\subsection{Mass Accretion Rates}\label{arevolsec}

Assuming that the bolometric luminosity in Class~0 and Class~I is due
entirely to accretion, we can estimate the accretion rate from \lbol:
\begin{equation} \label{dmdteq}
\dot{M} = \frac{dM}{dt} \sim \frac{2 R_* L_{bol}}{G M_*},
\end{equation}
where $R_*$ and $M_*$ are the radius and mass of the embedded
protostar.  For the following we assume $M_* =0.5 \msun$ and $R_* = 5 R_{\sun}$, 
but accretion rates can be easily scaled for different values of the mass or radius.
A mass of $0.5 \msun$ is consistent with the initial mass function \citep[e.g][]{chab03} 
and with average YSO masses \citep{merin08,spezzi08}; a radius of $3-5 R_{\sun}$ is typical 
for pre-main sequence models of low mass sources \citep[e.g][]{ps91,rob06}.

For comparison the ``Shu accretion rate'' of $c_s^3/G$, which is used
in the YE05 models, is approximately $4\times10^{-6}$ \msun\ yr$^{-1}$
for $c_s = 0.2$ km s$^{-1}$.  The M98 model begins with $\dot{M} \sim
10^{-6}$ \msun\ yr$^{-1}$, falling to $10^{-9}$ \msun\ yr$^{-1}$ by the
time it reaches the main sequence.   Making a $1~ \msun$ star in
$5.4\times 10^5$ yr requires an average accretion rate of approximately
$2 \times 10^{-6}$ \msun\ yr$^{-1}$.

\begin{figure*}[!ht]
\vspace{-0.6in}
\hspace{-0.3in}
\includegraphics[width=7.3in]{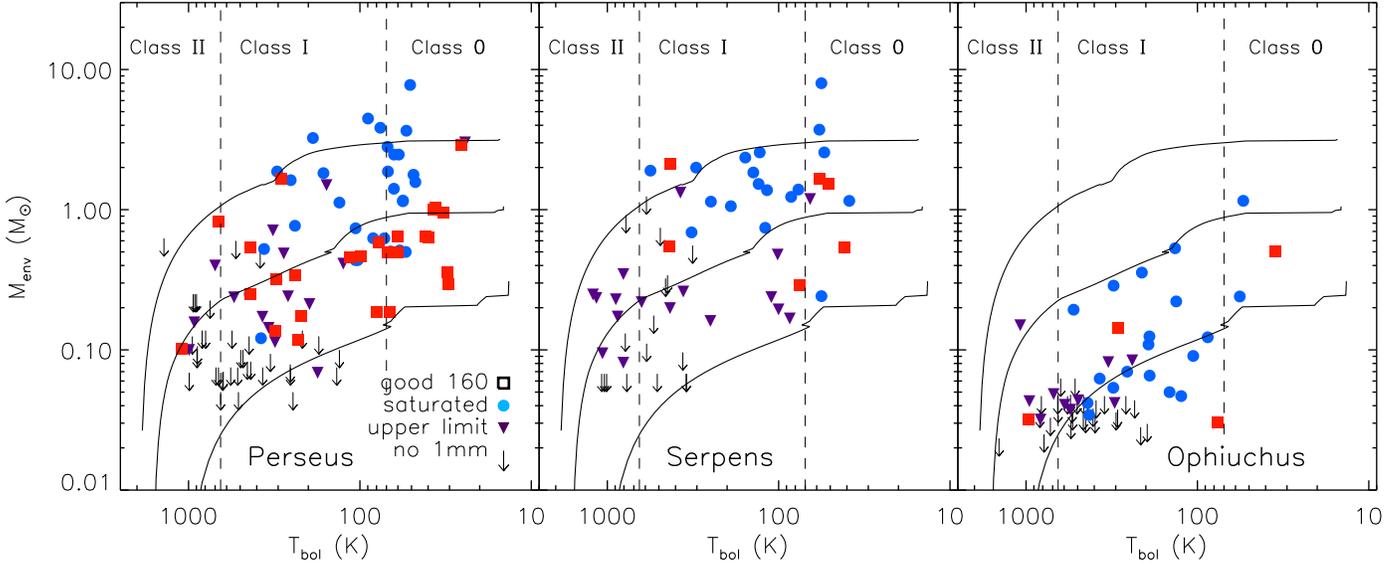} 
\vspace{-0.3in}
\caption
{Envelope mass versus bolometric temperature for embedded protostars
  in Perseus, Serpens, and Ophiuchus.  Filled and open symbols
  indicate that a given protostar is associated with 1.1~mm emission,
  while sources with upper limits at 1.1~mm are plotted as arrows.
  Solid lines show the predictions of protostellar evolutionary models
  from \citet{ye05} for sources of mass (from top to bottom) 3.0, 1.0,
  0.3~$M_{\sun}$.  The \citet{ye05} models describe the evolution of
  envelope mass with $T_{bol}$ quite well,  for sources of mass $M
  \sim 0.3 - 4 ~ \msun$ in Perseus and Serpens, and $M \sim 0.1-1.5 ~
  \msun$ in Ophiuchus.}
\label{bols2}
\end{figure*}

The mean luminosity of Class~I sources corresponds to $\dot{M}
\sim 1-2 \times 10^{-6}$ \msun\ yr$^{-1}$, not far from the average
required to make a solar mass star.   By contrast, the luminosity of
the low-\lbol\ ($L_{bol} \sim0.1 \lsun$) Class~I sources implies a
mass accretion rate of only $7\times 10^{-8}$ \msun\ yr$^{-1}$,
substantially smaller than the Shu value or the average required to
make a solar mass star.  This result again suggests that either these
sources will form very low mass stars ($M_* \lesssim 0.05
\msun$)\footnote{Although this would require these sources
  to have a quite low core-to-star formation efficiency of $\lesssim
  10\%$, as only a few Class~0 sources have envelope masses less than
  0.5\msun.},  or that these sources are in a suppressed accretion
phase of an episodic cycle.   If we define ``sub-Shu'' accretion to be
$\dot{M} \lesssim 10^{-6}$ \msun\ yr$^{-1}$, or $\lbol \lesssim 1
\lsun$, then approximately 55\% of Class~I sources are observed to be in
such a ``sub-Shu'' accretion state.

If the episodic accretion picture is correct there must be a
population of sources with accretion rates much higher than the
average; indeed, the highest observed luminosities ($\lbol \sim 10-30
~\lsun$) imply $\dot{M} \sim 1-2\times 10^{-5}$ \msun\ yr$^{-1}$, at least a factor of five
higher than the Shu value or the average required to make a
$1\msun$ star.  Approximately 5\% of the Class~I sources appear to be 
in such a ``super-Shu'' accretion rate ($\lbol > 10 \lsun$, or 
$\dot{M} \gtrsim 10^{-5}$ \msun\ yr$^{-1}$.

Unfortunately, it is very difficult to determine the duty cycle of the
episodic accretion without detailed star formation and accretion
models.  The small fraction of sources observed to have very high
accretion rates (5\% with $\dot{M} \gtrsim 10^{-5}$ \msun\ yr$^{-1}$),
suggests that periods of rapid accretion must be fairly short lived.
\citet{evans08} employ a simple accretion model to estimate the time
spent in accretion and quiescent phases, finding that half the mass of
a 0.5\msun\ star could be accreted during 7\% of the Class~I lifetime.

\subsection{Envelope Mass Evolution}\label{mevolsec}

Protostellar evolution models also predict the evolution of envelope
mass, $M_{env}$, with $T_{bol}$, as shown in Figure~\ref{bols2}.
Symbols are similar  to Figure~\ref{bols1}, with upper limits at
1.1~mm  represented by arrows, and solid lines indicating the YE05
evolutionary tracks (3.0, 1.0, and 0.3~\msun\ top to bottom).   Some of
the higher \tbol\ sources have high 1.1~mm upper limits because they
lie in regions of extended or confused emission.  Our determination
that these sources are not actually associated with the 1.1~mm
emission is based on visual examination of the images and SEDs.

Sources in all three clouds show a weak but consistent trend of
decreasing $M_{env}$ with increasing $T_{bol}$, as expected if the
envelope is gradually depleted by accretion onto the protostar.  This
trend is not a result of including the 1.1~mm flux, from which the
envelope mass is derived, in the calculation of $T_{bol}$; the same
trend is apparent even if the 1.1~mm point is excluded in the
calculation of $T_{bol}$.

Here, the YE05 model tracks fit the observed $M_{env}-T_{bol}$
distribution quite well.  Thus, a constant envelope infall rate reproduces
the decrease in envelope mass with increasing $T_{bol}$, although it
does not fit the evolution of $L_{bol}$ with $T_{bol}$
(Figure~\ref{bols1}).   

For standard episodic accretion models, where infall from the envelope 
is steady and the accretion rate from the disk to the protostar is 
the variable quantity \citep[e.g.,][]{kh95}, the envelope mass will
steadily decrease with increasing $T_{bol}$ even for variable
accretion.   Note that accretion stops when the envelope mass, as
defined  by the outer radius, has been exhausted.  In the context of
the YE05 models, the spread of $M_{env}$ as a function of $T_{bol}$
suggests stellar masses in the range $M \sim 0.3 - 4 ~ \msun$ in
Perseus and Serpens, and $M \sim 0.1-1.5 ~ \msun$ in Ophiuchus.

\section{Lifetime of the Class~0 Phase}\label{lifesec}

The length of time that sources spend in the Class~0 phase is an
important  diagnostic of protostellar evolution and how accretion
rates evolve with time.   If the rate of star formation in these
clouds is steady in time (i.e., not occurring in bursts), and if
Class~0--Class~I--Class~II  represents a true evolutionary sequence,
then we can use the number of objects in consecutive evolutionary
phases to estimate the relative lifetimes of those phases: $t_1/t_2 =
N_1/N_2$.    As we ultimately calibrate lifetimes based on the
lifetime of  Class~II disks ($\sim2 \times 10^6$~yr;
\citealt{ken90,cieza06,spezzi08}), star formation must have been
steady in time for at least the last 2~Myr.  In addition, there can be
no significant dependence of the evolutionary timescales on source
mass.

It is unlikely that all of these assumptions hold in every star
forming region.  In fact there is some observational evidence for
mass-dependent evolution of dense cores \citep{hatch08}, which could
easily translate into  mass-dependent evolution after protostar
formation.  We can, however, hope to mitigate the effects of any
breakdowns in our assumptions by utilizing our large sample and
averaging over three different environments. 

In a companion paper \citep{enoch08} we use a similar argument to
derive the lifetime of the prestellar phase from the ratio of the
number of starless cores to the total number of embedded protostars
(Class~0 + Class~I).  There we find there that the lifetime of dense
($n \gtrsim 10^4$ cm$^{-3}$) prestellar cores is approximately equal
to the lifetime of the embedded protostellar phase, or
$2-5\times10^5$~yr, suggesting that such cores evolve dynamically over
a few free-fall timescales.

The relative number of Class~0 and Class~I sources in each cloud,  and
the Class~0 lifetime derived from that ratio, is given in
Table~\ref{lifetab}.   Given in parentheses are
the values resulting from using the prismoidal, rather than midpoint
method for determining \tbol;  the Class~0 lifetime is based on an
average of the two methods, and the difference between them, added in
quadrature with $\sqrt{N}$ statistical uncertainties, yields an
uncertainty for the lifetime.

There are approximately half as many Class~0 as Class~I sources in
both Perseus and Serpens ($N_{\mathrm{Class~0}}/N_{\mathrm{Class~I}} = 0.7$ and 0.4,
respectively),  suggesting that the Class~0 phase lasts roughly half
as long as the  Class~I phase.  We adopt a total embedded phase lifetime
of $t_{emb} = t_{Cl 0} + t_{Cl I} \sim 5.4\times 10^5$~yr, 
derived based on the relative number of embedded protostars and 
Class~II sources \citep{evans08}.
Thus our measured ratios imply Class~0
lifetimes of $t_{\mathrm{Class~0}} \sim 2.2 \times 10^5$~yr in Perseus,
and $1.7 \times 10^5$~yr in Serpens.    In Ophiuchus there are
only 3 or 4 Class~0 sources, resulting in a ratio of 
$N_{\mathrm{Class~0}}/N_{\mathrm{Class~I}} = 0.1-0.2$ and
$t_{\mathrm{Class~0}} \sim 0.7 \times 10^5$~yr. 

Taking all three clouds together, we find a lifetime for the Class~0 
phase of $1.72\pm0.25\times10^5$ yr. 
This value is significantly longer than a number of previous estimates of
$t_{\mathrm{Class~0}} \sim 10^4$~yr, based both on the number of Class~0 sources
in Ophiuchus \citep{am94}, and on comparison to evolutionary models
($2-6\times10^4$~yr; \citealt{fro06}).  A short Class~0 lifetime is generally
interpreted as evidence for a period of very rapid accretion
early on in the evolution of protostars, causing them to quickly reach Stage~I, 
at which point the accretion rate decreases significantly. 
Our results argue against such a rapid accretion
phase.  Although accretion may decrease somewhat in Class~I
(or become episodic, see \S\ref{evolsec}), it appears unlikely that
the average accretion rate drops by more than a factor of two from 
Class~0 to Class~I, based on the relative lifetimes of the two phases.  
Similar mean luminosities for the Class~0 and Class~I phases 
also argue against very high accretion rates in Class~0.

Our derived Class~0 lifetime is similar to 
the results of \citet{vrc02} for a sample of Lynds dark clouds
($t_{\mathrm{Class~0}} \sim 2\times 10^5$~yr), and to the recent findings
of \citet{hatch07} that the Class~0 lifetime in Perseus is similar to
the Class~I lifetime ($2.5-6.7 \times 10^5$~yr with 95\% confidence).

Our large, unbiased sample provides a distinct advantage over many
other previous studies, which have necessarily relied on small samples
or accumulated sources from a number of different surveys,
wavelengths, and detection methods.   For example, the \citet{vrc02}
Class~0 sample consists of 7 sources, and the \citet{am94} lifetime
for Ophiuchus is based on one Class~0 object.
\citet{fro06} note that their 50 Class~0/I sources
are selected from a variety of surveys including NIR imaging of outflows,
IRAS data, submillimeter  and millimeter mapping, and radio continuum
surveys, causing their  source sample to be subject to strong
selection effects.   Within each cloud our
surveys are very uniform, providing protostellar samples that are 
envelope mass limited and not biased by selection effects.   
For this reason, our estimated Class~0
lifetimes should be more robust than most previous measurements.
The recent work by \citet{hatch07} comparing
SCUBA $850~ \micron$ maps and \textit{Spitzer} c2d data of Perseus, with
34 Class~0 sources, is a notable exception.  

Classifying sources based on $T'_{bol}$, calculated from photometry
corrected for extinction (see \S~\ref{deredsec}) results in a somewhat
smaller number of Class~0 sources, and a slightly shorter Class~0
lifetime.  Table~\ref{lifetabprime} gives the class statistics and
corresponding Class~0 lifetime derived using the extinction corrected
photometry.  The three-cloud average Class~0 lifetime is shorter than
when using observed photometry, by approximately 35\% ($t'_{\mathrm{class~0}}
= 1.1\times10^5$~yr).

\subsection{Cloud to Cloud Differences: Ophiuchus}\label{ophcl0life}
\begin{figure*}[!ht]
\includegraphics[width=6.5in]{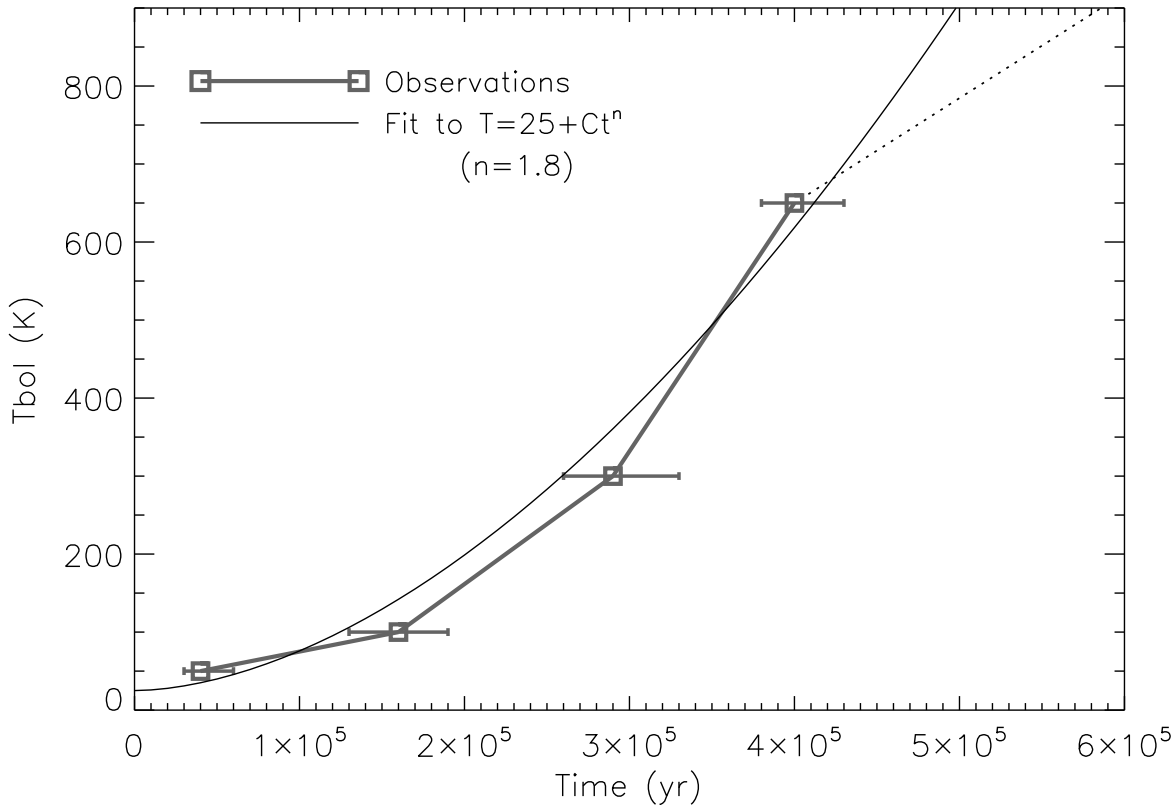}
\caption
{Evolution of the bolometric temperature with time, based on estimated
  lifetimes for four $T_{bol}$ intervals from $T_{bol}<50$~K to
  650~K.  $T_{bol}$ ranges, inferred lifetimes, and uncertainties are
  taken from Table~\ref{lifetab2}, and plotted as thick gray lines and
  symbols.  The dashed line connects to the Class~II lifetime at
  $t = 2\times10^6$ yr and $T_{bol}=2800$ K.  We fit a simple function to
  $T_{bol}$ versus time: $T_{bol} = 25~ \mathrm{K}+C (t/10^5\mathrm{~yr})^n$; the best fit
  is for $T_{bol} \propto t^{1.8}$.}
\label{timetbol}
\end{figure*}

The results in Ophiuchus are strikingly different from  Perseus and
Serpens, with nearly 10 times more Class~I than Class~0 sources.  
Our results for Ophiuchus are similar to previously observed
ratios of $N_{\mathrm{Class~0}}/N_{\mathrm{Class~I}} \sim 1/10$ in that cloud
\citep{am94}.
The derived Class~0 lifetime, $t_{\mathrm{Class~0}} \sim 7 \times 10^4$~yr, is 
still substantially longer than  the very short $t_{\mathrm{Class~0}} \sim 10^4$~yr 
found by \citet{am94}, however.   

There are two obvious, but conflicting, possible explanations for the 
smaller $N_{\mathrm{Class~0}}/N_{\mathrm{Class~I}}$ ratio in Ophiuchus.  First, the star
formation rate may be temporally variable in Ophiuchus.  \citet{vrc02}
suggest that a burst of star formation approximately $10^5$~yr ago is
responsible for the large ratio of Class~I to Class~0 sources.
Alternatively, the Class~0 phase  may be much shorter in Ophiuchus
than in the other two clouds due to higher accretion rates at early
times.  Were this the case, however, we would expect the mean
luminosity to be significantly higher for the Class~0 sources than for
the Class~I sources, which is not observed.  While a burst of star
formation seems the more likely explanation, accretion rates, and thus
lifetimes, could conceivably depend on cloud environmental factors
such as mean density or turbulent velocity.

On the other hand, the observed number of Class~0 sources in Ophiuchus
may be biased by the dearth of $160~ \micron$ flux measurements in that
cloud.  As discussed in \S\ref{appendsect}, bolometric temperatures
are likely overestimated for sources without a $160~\micron$ fluxes,
which may cause up to 50\% of Class~0 sources to be classified as
Class~I. The lack of $160~ \micron$ fluxes is especially problematic
in Ophiuchus, where the majority of sources are either saturated at
$160~ \micron$ or in regions of poor coverage, and may be partially
responsible for the small number of observed Class~0 sources.

\subsection{Limitations on the Class~I Sample}\label{limitsec}

Recall that we only include sources in our Class~I sample if they are
detected at 1.1~mm.  This criteria excludes 24 sources from the
original candidate protostar sample in Perseus with $70 < T_{bol} <
650$~K, 18 sources in Serpens, and 36 in Ophiuchus.  The average
spectrum of these ``non-envelope Class I'' sources is shown in
Figure~\ref{avespec3} (Perseus and Serpens only), and appears to be
intermediate between the Class~I  and Class~II averages. 
Furthermore, the mean $T_{bol}$ values of the non-envelope Class Is
(570~K in Perseus, 520~K in Serpens, and 500~K in Ophiuchus) confirm
that they fall at the warm end of the Class~I distribution.   While it
is possible that we are simply not detecting the envelopes of these
sources (e.g., if $M_{env} \lesssim 0.1 ~ \msun$ or they are very
diffuse),  we suggest that the transition from Stage~I (protostars
that retain an envelope) to Stage~II (pre-main sequence stars with no
envelope) occurs closer to $T_{bol}=400-500$~K than the standard
$T_{bol} = 650$~K, for these data.

When we add the ``non-envelope Class~I'' sources to our Class~I
samples, dropping the requirement that they be detected at 1.1~mm, the
Class~0 to Class~I ratio becomes 0.5 in Perseus, 0.3 in Serpens, and
0.07 in Ophiuchus.  This represents  a 25--30\% decrease in the Class~0
lifetime for each cloud.  On the other hand, \citet{crapsi08} suggest
based on protostellar models that up to half of sources
observationally classified as Class~I may actually be Stage~II source
with disk inclinations $\gtrsim 65^{\mathrm{o}}$.  Our ``non-envelope
Class~I'' objects may be just such sources; if the \citet{crapsi08}
picture is correct, the observed number of Class~I sources likely
represents an \textit{upper} limit to the true number of Stage~I
protostars.

\subsection{Evolution of $T_{bol}$ with Time}\label{ttbolsec}

In Table~\ref{lifetab2} we give the number of sources in each cloud in
our narrow \tbol\ bins (``early Class 0'', ``late Class 0'', etc.; 
\S\ref{otherclasssec}).  The ratio of the total number in each bin, summed over
the three clouds, to the total number of embedded protostars is used
to derive a lifetime for each $T_{bol}$ range.  The mean $T_{bol}$
within that range is also given. 

With the exception of ``early Class 0'', there is no evidence for a
difference in the number of sources in each \tbol\ bin, and little
difference in the derived lifetimes.  Notably, 75\% of the embedded
phase lifetime has elapsed by the time the bolometric temperature
reaches 300 K, and the mean $T_{bol}$ in ``Late Class~I'' (400 K) is
skewed toward the lower temperature end of the bin.   These features
support our earlier suggestion that the dividing line between Stage~I
and Stage~II is probably closer to $T_{bol}=400-500$~K than to 650 K.
$T_{bol} = 50$~K is reached at quite early times, which is not 
unexpected if the temperature scale starts at approximately 10 K with 
starless cores.  

With lifetime measurements in several $T_{bol}$ intervals, together
with our previous conclusion that the average accretion rate is
approximately constant through the embedded phase, we can empirically
characterize the evolution of $T_{bol}$ with time.
Figure~\ref{timetbol} plots the evolution of the bolometric
temperature with time, based on values from Table~\ref{lifetab2}.  The
relationship is clearly non-linear; given the small number of observed
points, we fit a simple function to $T_{bol}$ as a function of time:
\begin{equation}
T_{bol} = 25~ \mathrm{K}+C \left(\frac{t}{10^5\mathrm{~yr}}\right)^n,
\end{equation}
where C is constant and 
we choose 25~K as the temperature at $t=0$ because that is the minimum observed
$T_{bol}$ in Class~0.  The best fit is for C~$=51$~K and $T_{bol} \propto t^{1.8}$,
indicating a fairly steep evolution with time.  The dependence on time
must flatten out significantly after the embedded phase, however, to
match the lifetime for Class~II (dashed line in Figure~\ref{timetbol}).

\section{Conclusions}\label{discsect}

Utilizing large-scale 1.1~mm surveys \citep{enoch06,young06,enoch07}
together with \textit{Spitzer} IRAC and MIPS maps from the c2d Legacy
program \citep{evans03}, we have constructed an unbiased census of
deeply embedded protostars in the Perseus,
Serpens, and Ophiuchus molecular clouds.  
Our sample includes a total of
39 Class~0 sources and 89 Class~I sources, with approximate detection
limits of $M_{env} \gtrsim 0.1~ \msun$ and $\lbol \gtrsim 0.05 \lsun$ 
for the envelope mass and bolometric luminosity, respectively.
We also detect a few Class~II and Herbig Ae/Be candidates at 1.1~mm, 
most likely evidence of fairly massive proto-planetary disks.

Bolometric luminosities, temperatures, and envelope masses are
calculated for the candidate Class~0 and Class~I sources in each
cloud.  We compare protostellar classification methods,
concluding that, for deeply embedded sources, the bolometric
temperature \tbol\ is a better measure of evolutionary state than the
near- to mid-infrared spectral index, $\alpha_{IR}$.  

We also explore
classifying sources into ``early Class 0'', ``late Class I'',
etc., based on dividing them into narrower \tbol\ bins.
Average observed spectra in these bins are compared to model
predictions from \citet{whit03} for ``early Stage 0'', ``late Stage
0'', etc. 
In a broad sense the Stage~0 and Stage~II models match the observed 
spectra fairly well.  The agreement with both Stage~I models is
quite poor, however, as the models severely over-predict the near-IR 
and under-predict the mid-IR flux at this stage, displaying a 
double-peaked SED that is rarely observed.

Observed source properties are compared to protostellar evolutionary
models using the bolometric temperature-luminosity ($L_{bol}-T_{bol}$)
diagram, the protostellar equivalent of the H-R diagram
\citep{myer98}.  Neither models with a constant mass accretion rate
\citep{ye05}, nor those with an exponentially decreasing rate
\citep{myer98} fit the observed sources.  

In particular, there is a large population of low luminosity Class~I
sources that  aggravate the previously noted ``luminosity problem''
for embedded protostars \citep[e.g.][]{ken90}.  We interpret this
result as evidence for episodic accretion beginning at least by the
Class~I  phase, and possibly earlier.  More than 50\% of Class~I
sources are inferred to have ``sub-Shu'' mass accretion rates
($\dot{M} \lesssim 10^{-6}$ \msun\ yr$^{-1}$, corresponding to $\lbol
\lesssim 1 \lsun$), and approximately 20\% have $\dot{M} \lesssim
10^{-7}$ \msun\ yr$^{-1}$.  To build up of order a solar mass in
$5.4\times 10^5$~yr, such sources must also have periods of
``super-Shu'' accretion ($\dot{M} \gtrsim 10^{-5}$ \lsun yr$^{-1}$).
Few very high luminosity sources are observed (5\%), suggesting that
such rapid accretion periods must be short lived. An important caveat
to this analysis is that we may sometimes underestimate  the
luminosity due to missing 70 or 160~$\micron$ photometry. 

Finally, the relative number of Class~0 and Class~I sources are used
to estimate the lifetime of the  Class~0 phases.  There are
approximately half as many Class~0 as Class~I sources in the 
three cloud sample, implying an average Class~0 lifetime of 
$1.7\pm0.3\times10^5$~yr ($1.1\times 10^5$ yr when approximate 
extinction corrections are applied).  This
lifetime rules out drastic changes  in the mass accretion rate from
Class 0 to Class I, particularly extremely rapid accretion in Class~0.  
In Ophiuchus the fraction of Class~0 sources is
much smaller.  While this difference could be due to the lack of $160~ \micron$ flux
measurements in Ophiuchus, it may be that either the Class~0 phase
is shorter in that cloud ($0.7\times10^5$~yr), or that a burst of star formation
is responsible for the large population of Class~I objects
\citep[e.g. as suggested by][]{vrc02}.  

Altogether, the large variation in Class~I luminosities, similar mean
luminosities for the Class~0 and Class~I phases, and not dramatically
different numbers of Class~0 and Class~I sources (at least in Perseus
and Serpens) suggests a consistent picture of nearly constant
\textit{average} mass accretion rate through the entire embedded phase
(Stage~0 and Stage~I), with highly variable episodic accretion turning
on by at least early Stage~I, and possibly sooner.  Understanding
embedded protostellar structure and evolution well enough to reproduce the
observations with detailed models of spectra and evolution presents an
ongoing challenge.

\acknowledgments

The authors thank J. Hatchell, Y. Shirley, and the anonymous referee for comments and 
suggestions that helped to improve this paper, as well as M. Dunham for many fruitful discussions.
We are grateful to B. Whitney for sharing the protostellar evolutionary model data used here. 
We thank the Lorentz Center in Leiden for hosting meetings that contributed to this paper.  
Part of the work was done while in residence at the Kavli Institute for Theoretical Physics 
in Santa Barbara, California.
Support for this work, part of the Spitzer Legacy Science Program, was
provided by NASA through contracts 1224608 and 1230782 issued by the
Jet Propulsion Laboratory, California Institute of Technology, under
NASA contract 1407.  Additional support was provided by NASA 
through the Spitzer Space Telescope Fellowship Program and obtained from NASA Origins
Grants NNG04GG24G and NNX07AJ72G to the University of Texas at Austin.  Support for
the development of Bolocam was provided by NSF grants AST-9980846 and
AST-0206158.

\clearpage

\vspace{2.5in}

\begin{deluxetable*}{ll}
\vspace{1in}
\tablecolumns{2}
\tablewidth{0pc}
\tablecaption{\label{deftab}Definition of Classes and Stages}
\tablehead{
\colhead{Class/Stage}  & \colhead{definition}}
\startdata
Class 0 & $T_{bol} \le 70$~K \\
Class I & 70 K$< T_{bol} \le 650$~K; 1.1~mm detection ($M_{env} \gtrsim 0.1 M_{\sun}$) \\
Class II & 650 K$<T_{bol} \le 2800$~K \\
\hline
~~Early Class 0 & $T_{bol} \le 50$~K \\
~~Late Class 0 & 50 K $<T_{bol} \le 100$~K \\
~~Early Class I & 100 K $<T_{bol} \le 300$~K \\
~~Late Class I & 300 K $<T_{bol} \le 650$~K \\
~~non-envelope Class~I & 70 K$< T_{bol} \le 650$~K; \textit{no} 1.1~mm detection ($M_{env} \lesssim 0.1 M_{\sun}$)\\
\hline
Stage 0  &  $M_{*} < M_{env}$ \\
Stage I  &  $M_{*} > M_{env}$; $M_{env} \ge 0.1 M_{\sun}$ \\
Stage II &  circumstellar disk; $M_{env} < 0.1 M_{\sun}$ 
\enddata
\end{deluxetable*}
\vspace{1in} 

.\\
\input{tab2}
\input{tab3}
\input{tab4}

\begin{deluxetable*}{lcccc}[!ht]
\tablecolumns{5}
\tablewidth{0pc}
\tablecaption{\label{lifetab}Numbers of Protostars by Class and the Class~0 lifetime}
\tablehead{
\colhead{Cloud}  & \colhead{$\mathrm{N_{Class 0}}$ \ \ } & \colhead{\ \ $\mathrm{N_{Class I}}$ \ \ } & \colhead{ \ \ $\mathrm{N_{Class 0}/N_{Class I}}$ \ \ } & \colhead{$\tau_{Class 0}$}  \\
\colhead{} & \colhead{} & \colhead{} & \colhead{} & \colhead{(yr)} 
}
\startdata
Perseus   &   27 (27) &  39 (41) &  0.70 (0.66) & \ \  $2.2\pm0.4\times10^5$  \ \  \\ 
\vspace{0.1cm}
Serpens   &   9 (13)  &  25 (23)  &  0.4 (0.6) & \ \  $1.7^{+0.5}_{-0.9}\times10^5$  \ \   \\ 
\vspace{0.1cm}
Ophiuchus \ \ &   3 (4)  &  25 (24) &  0.12 (0.17) & \ \   $0.7^{+0.4}_{-0.5}\times10^5$ \ \    \\
Three cloud sample  &  39 (44)  &  89 (88)  &  0.44 (0.50)  &  $1.72 \pm 0.25\times10^5$ \\
\enddata
\tablecomments{Numbers of Class~0 and Class~I sources are based on $T_{bol}$ classifications (Table~\ref{deftab}) including the requirement that Class~I sources be detected at 1.1~mm.  Numbers in parentheses indicate how the results change if we utilize a different method for calculating $T_{bol}$ (prismoidal versus midpoint integration, see \S\ref{appendsect}).   The Class 0 lifetime, $\tau_{Class 0}$, assumes that the entire embedded phase lasts for $5.4\times 10^5$ years ($\tau_{emb} = \tau_{Class 0} + \tau_{Class I} = 5.4\times 10^5$~yr; \citealt{evans08}), and $N_{Class 0}/(N_{Class 0} + N_{Class I}) = \tau_{Class 0}/\tau_{emb}$.  The lifetime is calculated using the average of the two $N_{Class 0}/N_{Class I}$ ratios given, and the uncertainty is from the difference between them added in quadrature with the $\sqrt{N}$ uncertainties from counting statistic. 
}
\end{deluxetable*}

\input{tab5}

\begin{deluxetable}{lcccc}[!ht]
\tablecolumns{5}
\tablewidth{0pc}
\tablecaption{\label{lifetabprime}Effect of Extinction Corrections on Class Statistics}
\tablehead{
\colhead{Cloud}  & \colhead{$\mathrm{N_{Class 0}}$ \ \ } & \colhead{\ \ $\mathrm{N_{Class I}}$ \ \ } & \colhead{ \ \ $\mathrm{N_{Class 0}/N_{Class I}}$ \ \ } & \colhead{$\tau_{Class 0}$}  \\
\colhead{} & \colhead{} & \colhead{} & \colhead{} & \colhead{(yr)} 
}
\startdata
Perseus   &   21 &  42 &  0.5 & \ \  $1.5\times10^5$  \ \  \\ 
Serpens   &   4   &  26   &  0.15 & \ \  $0.6\times10^5$  \ \   \\ 
Ophiuchus \ \ &   3  &  20 &  0.15 & \ \   $0.6\times10^5$ \ \    \\

Three cloud sample  &  29  &  88  &  0.33  &  $1.1\times10^5$ \\
\enddata
\tablecomments{Same as Table~\ref{lifetab} but for classifications based on $T'_{bol}$, the extinction-corrected bolometric temperature. In addition, an embedded phase lifetime of $4.4\times 10^5$~yr is assumed, derived based on extinction-corrected classifications \citep{evans08}.}
\end{deluxetable}

\begin{deluxetable}{llcccccc}[!ht]
\tabletypesize{\footnotesize}
\tablecolumns{8}
\tablewidth{0pc}
\tablecaption{\label{lifetab2}Lifetimes for narrow $T_{bol}$ bins}
\tablehead{
\colhead{Phase} & \colhead{$T_{bol}$ range}  & \colhead{N(Per)} & \colhead{N(Ser)} & \colhead{N(Oph)} & \colhead{N(total)/} & \colhead{Mean $T_{bol}$} & \colhead{Lifetime} \\
\colhead{ } & \colhead{(K)}  & \colhead{} & \colhead{} & \colhead{} & \colhead{N(emb,tot)} & \colhead{(K)} & \colhead{(yr)} 
}
\startdata
\vspace{0.1cm}
Early Class~0   & $T < 50$        &  11  &  2  &  1  &  $14/128 = 0.11$  &  40  &  $0.6\times10^5$ \\
\vspace{0.1cm}
Late Class~0    & $50 < T < 100$  &  23  &  11 &  4  &  $38/128 = 0.30$  &  70  &  $1.6\times10^5$ \\
\vspace{0.1cm}
Early Class~I   & $100 < T < 300$ &  19  &  12 &  12 &  $43/128 = 0.34$  &  180  &  $1.8\times10^5$ \\
\vspace{0.1cm}
Late Class~I    & $300 < T < 650$ &  13  &  9  &  11 &  $33/128 = 0.26$  &  400  &  $1.4\times10^5$
\enddata
\tablecomments{For each $T_{bol}$ bin, N(total) is the total number of sources in all three clouds; N(emb,tot) is the total number of embedded protostars in the three cloud sample (128).  Only sources detected at 1.1~mm are included.  The mean $T_{bol}$ of protostars in a given $T_{bol}$ range is also given.  Lifetimes are calculated from $\mathrm{N(total)/N(emb,tot)} \times 5.4\times10^5$ yr.  Cloud-to-cloud variations are typically $0.2-0.3\times 10^5$ yr.}
\end{deluxetable}

\clearpage

\appendix
\section{Calculating the bolometric luminosity and temperature}\label{appendsect}

\begin{figure*}[!hb]
\hspace{0.4in}
\includegraphics[width=6.in]{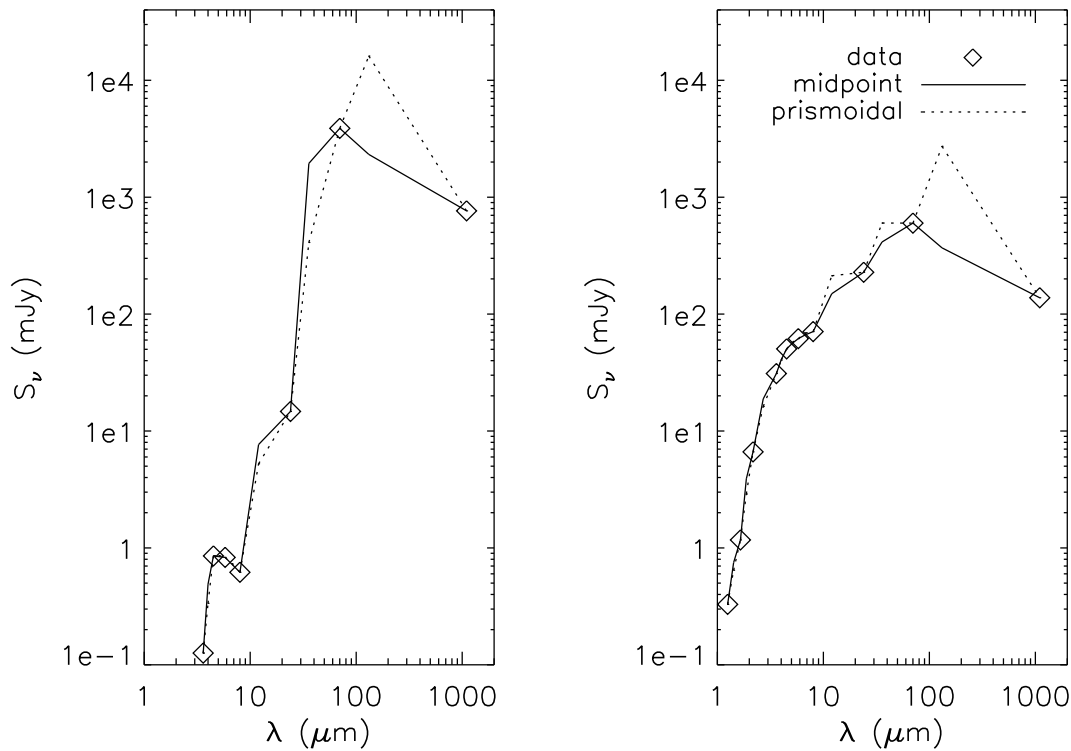}
\caption
{Examples of the interpolated midpoints used for the integration of
  the SED in the midpoint (solid line) and prismoidal (dashed line)
  methods.  Observed data points are overlaid as diamonds.  A typical
  Class~0 source is shown on the left, and a Class~I source on the
  right; both sources shown are lacking data at 160 $\mu$m.  After the
  midpoint is calculated, the integration for both methods is done in
  linear $\nu S_{\nu}$ space.  The prismoidal method provides a good
  approximation of the dust emission peak for cold
  sources (left), but probably overestimates the long-wavelength flux
  for warmer, flatter SED sources (right), resulting in a bias toward lower $T_{bol}$.} 
\label{demointeg}
\end{figure*}

Determination of the bolometric temperature and luminosity of any
given source can depend strongly on the method used to approximate the
integrations over frequency in
equations~\ref{lboleq}--\ref{meanfreqeq}.  As $S_{\nu}$ is sampled at
a finite number of frequencies (usually 6 to 10 here), the SED must be
interpolated over the intermediate frequencies.  Here we use two
different methods; the first (midpoint) method utilizes a simple
linear interpolation for the midpoint flux, while the second (prismoidal) method
calculates a color temperature for each pair of flux points,  and uses
a modified blackbody based on that color temperature to estimate the 
midpoint flux density.  In both cases the SED is extrapolated from the
longest wavelength flux using $S_{\nu} \propto \nu^2$, and  flux upper
limits are removed from the fit (i.e., we
interpolate over them).  Examples of the midpoint and prismoidal interpolations 
are shown in Figure~\ref{demointeg} for two sources with Class~0-like (left) and 
Class~I-like (right) SEDs. 

\begin{figure*}[!ht]
\includegraphics[width=7.2in]{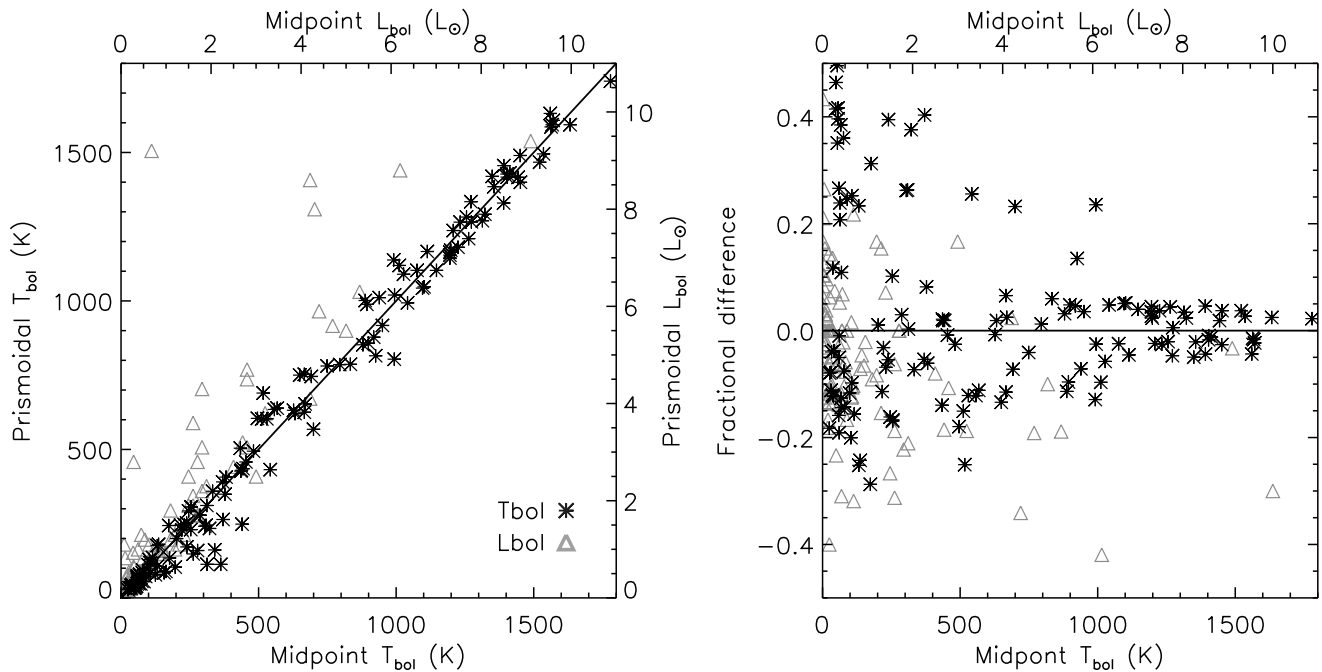}
\caption
{Comparison of the midpoint and prismoidal methods for calculating
  $L_{bol}$ and $T_{bol}$, for sources in Perseus.   \textit{Left}:
  Midpoint and prismoidal $T_{bol}$ (black asterisks) and $L_{bol}$
  (triangles) values are plotted for all candidate embedded protostars
  in Perseus.   \textit{Right}:  Fractional difference between the
  midpoint and prismoidal methods, plotted as
  (midpoint-prismoidal)/prismoidal.   The two methods generally agree
  to within 20\% for \tbol\, but $L_{bol}$ calculated by the
  prismoidal method is  higher than the midpoint $L_{bol}$ by as much
  as 100\%.   There is some tendency for the prismoidal $T_{bol}$ to be
  lower for colder sources.}
\label{tbolmeth}
\end{figure*}

Figure~\ref{tbolmeth} compares the results of using the midpoint and
prismoidal methods to calculate $L_{bol}$ and $T_{bol}$  for the
embedded protostar sample in Perseus.     The midpoint and prismoidal
$T_{bol}$ values agree fairly well, usually to within 20\%, with no
strong systematic bias.  There is some tendency for the prismoidal
$T_{bol}$ to be lower for colder sources.  On the other hand,
$L_{bol}$ calculated by the prismoidal method is  generally higher
than the midpoint $L_{bol}$.  

\begin{figure*}[!ht]
\includegraphics[width=7.2in]{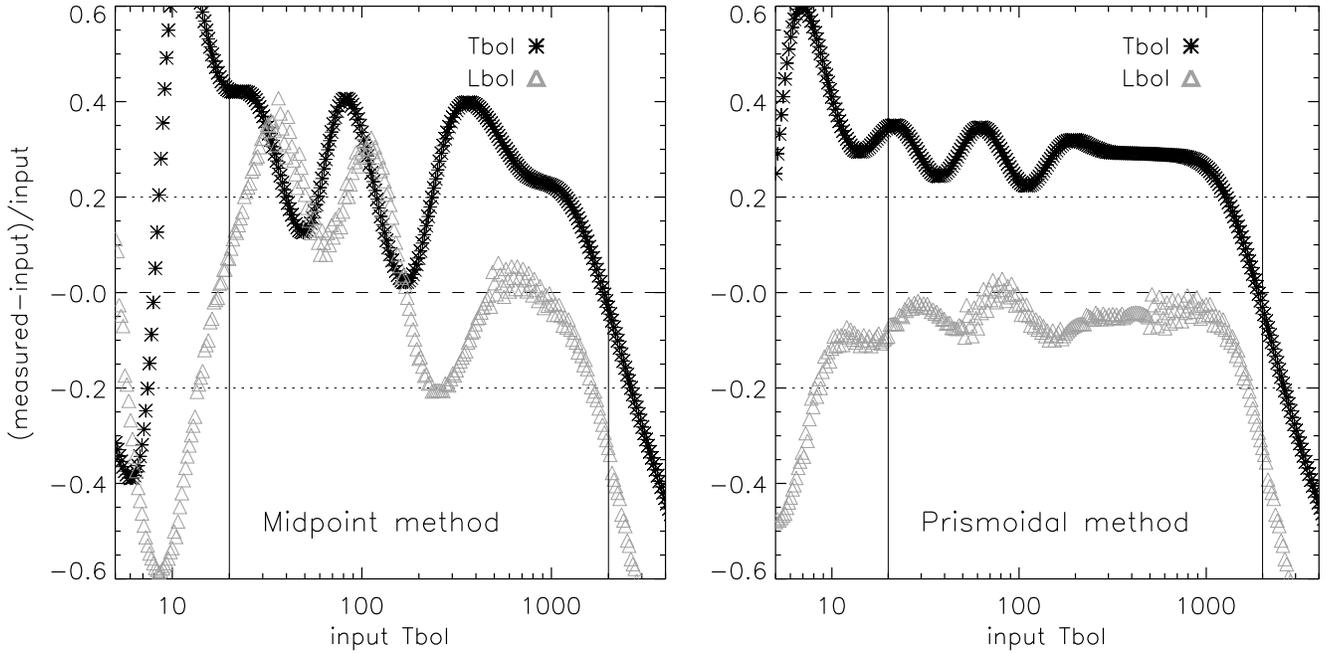}
\caption
{Characterization of sampling errors for $T_{bol}$ and $L_{bol}$ calculated
using the midpoint (left) and prismoidal (right) methods.  The
fractional difference between the input ($T_{bol}$,$L_{bol}$) of
blackbody spectra and the measured values when SEDs are sampled at the 
observed wavelengths is plotted as a function of the input $T_{bol}$.  
Both $L_{bol}$ and $T_{bol}$ have typical errors of approximately 20\%, 
and the measured $T_{bol}$ is consistently over-estimated by approximately 20\%.
Source SEDs, which are considerably more complex than the blackbodies tested here, may
behave differently.  }\label{tbolsamp}
\end{figure*}

To further test this behavior, we plot in Figure~\ref{tbolsamp} the
fractional errors due solely to finite sampling, for $L_{bol}$ and
$T_{bol}$  calculated from input blackbody spectra.  Blackbody sources
with input $T_{bol}$ from 5--5000~K are sampled at the observed
wavelengths ($\lambda=$1.25, 1.65, 2.17, 3.6, 4.5, 5.8, 8, 24, 70,
160, 1100 $\micron$), and the bolometric temperature and luminosity
calculated by the midpoint (left) and prismoidal (right) methods.   Both
$L_{bol}$ and $T_{bol}$ have typical errors of approximately 20\%,
and the measured $T_{bol}$ is consistently over-estimated by
approximately 20\%.  The midpoint \tbol\ is more variable but somewhat
more accurate than the prismoidal \tbol, while \lbol\ is somewhat less
accurate.  
Of course, the blackbody spectra tested here are not
necessarily representative of the more complicated observed SEDs.
Based on Figures~\ref{tbolmeth}--\ref{tbolsamp} and on examinations of
the midpoint and prismoidal fits to observed SEDs, from which we find
that the prismoidal method often provides a poor fit to ``flatter''
SEDs at long wavelengths (Figure~\ref{demointeg}), we use the midpoint 
method throughout this
paper.  Differences between the midpoint and prismoidal methods are
often used as a measure of uncertainty.

In general, measured $160~ \micron$ fluxes are uncertain by up to a
factor of 2, due to unquantified saturation effects and calibration
uncertainties, and in many cases we are unable to measure a
$160~ \micron$ flux at all.  In addition, the ``point-source''
$160~ \micron$ fluxes of extended sources will likely be underestimated,
because they are measured with a PSF fit rather than aperture
photometry.  Missing or severely underestimated fluxes at $160~ \micron$
can significantly affect both $L_{bol}$  and $T_{bol}$ for cold
sources that peak near $100~ \micron$.  As we are primarily interested
in the coldest sources, we attempt here to quantify the effects of
missing or underestimated $160~ \micron$ flux.  One way of approaching
this problem is to examine sources for which we have  additional
information near the peak of the SED, e.g., from IRAS or SHARC
$350~ \micron$ observations \citep{wu07}.

We first look at two example sources, one cold  (IRAS 03282+3039; 
$T_{bol}\sim 33$~K), and one warmer ($T_{bol} \sim 163$~K).   IRAS
03282+3039 is isolated, so we can use the IRAS fluxes without
worrying about confusion due to the large IRAS beam.  We calculate the
``true'' $L_{bol}$ and $T_{bol}$ by including the IRAS 60 and
$100~ \micron$ fluxes, a SHARC~II $350~ \micron$ flux, and a
MIPS $160~ \micron$ flux measured in a large aperture, which is almost
two times higher than the PSF-fit flux.  Using the underestimated
point source $160~ \micron$ flux, and no longer including IRAS or SHARC
II data, causes an \textit{underestimate} of $L_{bol}$ by 35\% and
an \textit{overestimate} of $T_{bol}$ by 15\% compared to the ``true''
values.  Omitting the $160~ \micron$ point altogether, as would be
appropriate for saturated sources, results in an underestimate of
$L_{bol}$ by 7\% and an overestimate of $T_{bol}$ by 9\%.  

Thus we conclude that while the $160~ \micron$ point is important for
characterization of the SED of embedded objects, our integration method
can interpolate over a missing $160~ \micron$ flux to  estimate
$L_{bol}$ and $T_{bol}$ to within 20\% for cold sources.  Severely
underestimated $160~ \micron$ flux densities will cause larger errors
of up to 50\%.  Using a similar procedure for the warmer source
($T_{bol}=163$~K), we find that omitting the $160~ \micron$ flux results
in errors in the opposite sense compared to the colder source:
an \textit{overestimate} of $L_{bol}$ by 28\% and
an \textit{underestimate} of $T_{bol}$ by 18\%.

\begin{figure*}
\plottwo{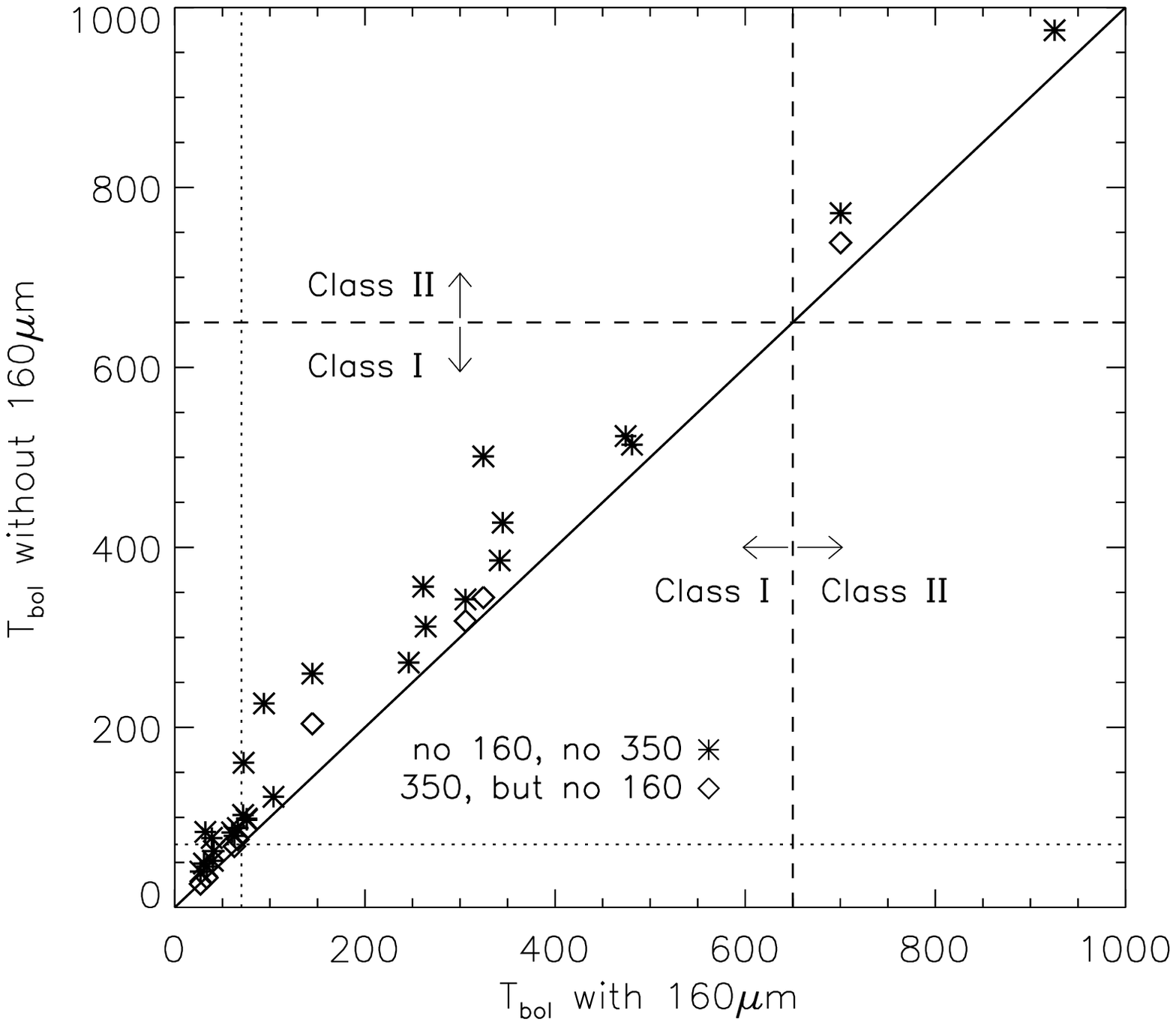}{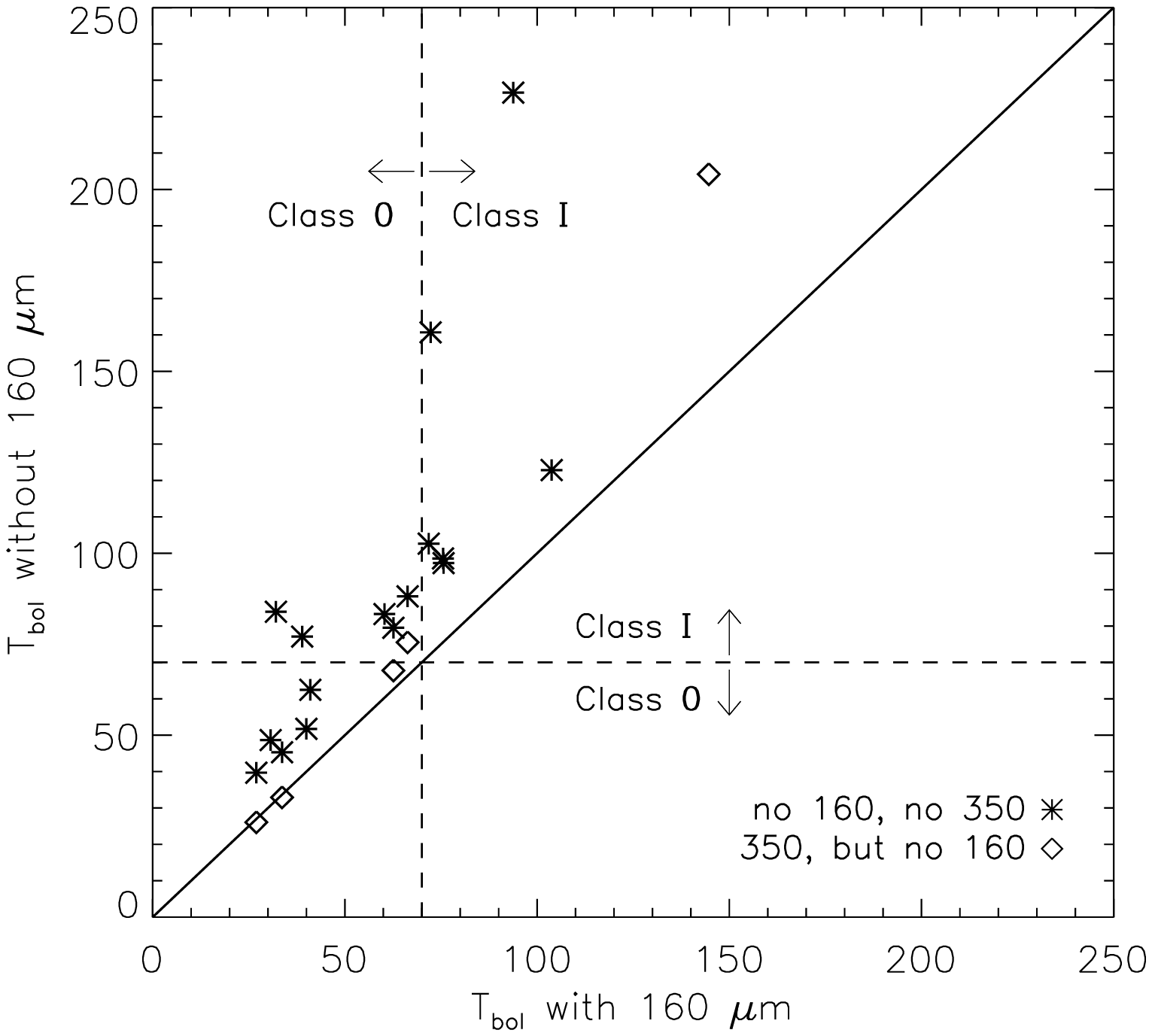}
\caption
{Characterization of the systematic errors introduced into the measured $T_{bol}$ 
when a $160~ \micron$ flux is not available.  \textit{Left}: For
all sources in Perseus with a reliable $160~ \micron$ flux, we calculate
$T_{bol}$ with and without the $160~ \micron$ point (asterisks).
$T_{bol}$ is always overestimated when the $160~ \micron$ flux is
missing, but it does not change the classification of Class~I sources
for these data. \textit{Right}:  Similar, but with the scale adjusted to 
highlight lower $T_{bol}$ sources.  Not having a $160~ \micron$ point does change the
classification of approximately half of the Class~0 ($T_{bol}<70$~K) sources.
Errors are substantially reduced when a flux at $350~ \micron$ is
available (diamonds).} 
\label{tbolerrs}
\end{figure*}

For a more general result, we compare $T_{bol}$ calculated with the
$160~ \micron$ flux included in the SED to the value found by omitting
the $160~ \micron$ point, for all sources in Perseus with a measured
$160~ \micron$ flux (Figure~\ref{tbolerrs}, left panel).   For this
plot, we have adopted classifications from \citet{chen95}, as discussed
in \S\ref{classsec}.  Although   $T_{bol}$ calculated without the
$160~ \micron$ flux is always overestimated compared to $T_{bol}$
calculated with the $160~ \micron$ flux, it is never a large enough
effect to shift the source classification  of objects with
$T_{bol} \gtrsim 100$~K (e.g., from Class~I to Class~II).   This is not
the case for Class~0 sources, however, as can be seen in the right
panel of Figure~\ref{tbolerrs}, where the scale has been adjusted to
highlight the lowest $T_{bol}$ values.  Five sources that have
$T_{bol} < 70$~K (Class~0) when including the $160~ \micron$ flux are
shifted to $650 < T_{bol} > 70$~K (Class~I) when the $160~ \micron$ flux is 
omitted.  If a published $350~ \micron$ flux is available (diamonds),
the errors in $T_{bol}$ resulting from excluding the $160~ \micron$ flux
are almost completely eliminated.  

Based on Figures~\ref{tbolmeth}--\ref{tbolerrs} and the above
examples, we estimate  overall uncertainties for measured $L_{bol}$ and
$T_{bol}$ values of $20-50$\%, depending on whether or not a $160~ \micron$
flux is available.   If we are unable to measure a $160~ \micron$ flux,
$T_{bol}$ will almost certainly be an overestimate for very cold
sources, which may affect our classification of the most deeply embedded
protostars.

\end{document}